\documentclass{amsart}
\RequirePackage{fix-cm}

\usepackage{wasysym}
\usepackage{amssymb,amsmath,stmaryrd,amsfonts}
\usepackage{mathtools,leftindex}
\usepackage{multirow}

\usepackage{siunitx}
\usepackage{array}
\usepackage{gensymb}
\usepackage{xcolor}
\usepackage{diagbox}
\usepackage{url}
\usepackage{bm}

\usepackage{tikz,mwe,pifont}
\tikzset{boximg/.style={remember picture,red,thick,draw,inner sep=0pt,outer sep=0pt}}

\usepackage{natbib}
\usepackage[margin=1.2in]{geometry}

\newcommand{\norm}[1]{\left\lVert#1\right\rVert}

\usepackage{enumitem}   

\usepackage{rotating}

\usepackage[colorlinks,citecolor=blue]{hyperref}
\newcommand{\ie}{\textit{i.e.,}\,}
\newcommand{\eg}{\textit{e.g.,}\,}
\newcommand{\vect}[1]{\ensuremath{\mathbf{#1}}}

\begin{document}

\title[$J_{2}$--solar radiation pressure problem]{Regularity and reentry basins of low Earth orbits in the $J_{2}$--solar radiation pressure problem}

\author[C.\,Barbis]{Cassandra Barbis}
\address{Nonlinear Dynamics and Chaos Group, Department of Mathematics and Applied Mathematics, University of Cape Town, Rondebosch,
Cape Town,
7701, South Africa}
\email{BRBCAS002@myuct.ac.za}

\author[J.\,Daquin]{J\'er\^ome Daquin}
\address{Univ Toulon, Aix Marseille Univ, CNRS, CPT, Toulon, France
}
\email{jerome.daquin@cpt.univ-mrs.fr}

\author[E.M.\,Alessi]{Elisa Maria Alessi}
\address{Instituto de Matematica Applicata e Tecnologie Informatiche, Consiglio Nazionale delle Ricerche
}
\email{elisamaria.alessi@cnr.it}

\author[C.\,Skokos]{Charalampos  Skokos}
\address{Nonlinear Dynamics and Chaos Group, Department of Mathematics and Applied Mathematics, University of Cape Town, Rondebosch,
	Cape Town,
	7701, South Africa
}
\email{haris.skokos@uct.ac.za; haris.skokos@gmail.com}

\date{\today}

\begin{abstract}
We numerically investigate the long-term dynamical structure of low Earth orbits (LEOs) using the Smaller Alignment Index (SALI), a fast numerical indicator of chaos, within a closed-form averaged model that incorporates the effects of solar radiation pressure and Earth's oblateness. Our analysis reveals that the area-to-mass ratio is a key parameter governing the onset and extent of chaotic behavior in LEOs. We map the system's chaotic regions, study the behavior of reentry trajectories and characterize their temporal laws over a timescale constrained by the $25$-year mitigation guideline. Within this physically relevant timescale, we show that most of the reentry trajectories exhibit regular motion. Reentry basins, constructed  according to different mitigation guidelines up to $25$ years, display fractal-like structures for less-stringent guidelines. The degree of this fractality is quantitatively assessed using the uncertainty exponent method. In most cases, for large area-to-mass ratios, reentry  occurs on relatively short timescales (a few years) -- short enough that no fractal behavior is observed in the basin boundaries. This numerical dynamical study offers insights into the development of dynamically informed deorbiting strategies.
\end{abstract}


\maketitle
\tableofcontents

\section{Introduction}\label{sec:intro}

Exploiting -- or ``surfing'' -- the natural flow properties of a dynamical system instead of counteracting its dynamics is observed in real-world, living systems and shows up across several engineering disciplines. A few examples include the use of hydrodynamics properties in planktonic and fish navigation \citep{rMo22,hKo23}, informed use of vortex dynamics in bio-inspired robotics for propulsion savings \citep{pGu25}, near-optimal path planning using Lagrangian coherent structures in ocean engineering \citep{agRa18}, and the increase or decrease of chaos in targeted mixing \citep{bTo06}. In the context of astrodynamics, examples include  the modeling of natural transport phenomena through invariant sets \citep{wKo00}, and the design of the Soviet Molniya constellation near the critical inclination to counteract the perigee drift \citep{aBu20}, to name only a few.  \\

Quite recently, it has been recognized that the use of natural disturbances and resonant phenomena might play a key role in managing the space environment to enable sustainable circumterrestrial operations, see, \eg  \cite{aRo16,aCe16,aRo18,dGo19,eLe23} and the works cited therein. 
Understanding the  effect of perturbations appears to be essential for end-of-life disposal (one of the most effective space debris mitigation measure) and monitoring of fragmentation events. Regarding the former,  depending on the orbital regime, several perturbations of the idealized two-body problem can be exploited, such as the Earth's non-spherical gravitational potential, the solar radiation pressure (SRP), the atmospheric drag, and the gravitational accelerations exerted by third bodies like the Moon and Sun. \\

This work focuses on the dynamical properties arising from the secular coupled effects of Earth’s oblateness and SRP, referred to hereafter as the $J_{2}$-SRP problem. A key aspect of this framework is the dependence of the dynamics on the area-to-mass ratio ($A/m$) of the orbiting object. This idealized model stems from  our desire to understand the behavior of a satellite equipped with a large solar sail deployed at the end of its operational life. The resulting high $A/m$, combined with carefully selected initial conditions near resonant configurations --  orbital inclination and orientation with respect to the Sun -- might enable the satellite to naturally reenter Earth’s atmosphere within a few years, without requiring active propulsion. This reentry is driven by a resonance-induced growth in orbital eccentricity \citep{aRo18, Alessi_MNRAS,ACR_CMDA2019,iGk20}.  Current technologies in solar sailing lead to an area-to-mass ratio of about $5\,\mathrm{m}^2\mathrm{kg}^{-1}$ \citep{Cetal18}. In the present study, we consider area-to-mass values up to $10\,\mathrm{m}^2\mathrm{kg}^{-1}$ 
to reflect large $A/m$ devices \citep{GEco63} and
account for possible future advancements \citep{REN2024234}. \\

Our work specifically differs from the recent works  \citep{Alessi_MNRAS, ACR_CMDA2019, iGk20} based on a single-resonance hypothesis, used to explain the so-called ``de-orbiting corridors'' numerically found  in \cite{ASRV_CMDA2018}. 
Here we extend the analysis of the $J_{2}$-SRP problem to investigate  whether the overall dynamical picture changes when \textit{all} resonant terms are included in the long-term model, with particular focus on large  $A/m$ devices.  
More specifically, we depart from the analytical endeavor and instead detect and portray numerically the location of chaos, its extent and geometry, associated timescale, and possible connection with reentry trajectories. 
The adopted $J_{2}$-SRP model provides a physically-motivated, reduced description of the long-term dynamics, especially for high   area-to-mass-ratio objects, whose resonant structures have been extensively analyzed and shown to persist in higher-fidelity models \citep{SASRS_ASR2019,cMa25,cMa26}. 
It does not pretend to cover dynamical features under a full model complexity; yet and
quite remarkably, this Hamiltonian construct has  been shown to support the observed resonant evolution of the historical Echo I satellite \citep{cMa26}.
Thus, in the regime and under the parameter sets investigated subsequently, the omission of drag is expected to have a limited impact on the dynamical structures investigated therein,  allowing us to isolate and characterize properties induced by the eccentricity growth mechanism without the sophistication of higher-order geopotential or lunisolar perturbations. In this sense, this model once again provides a worthwhile testbed constituting a first step toward the broader goal of understanding the structures studied here in more comprehensive orbital models.
\\

The complexity of the lifetime problem is typically discussed  in the context of reentry date prediction and event location, due to uncertainties in atmospheric drag modeling \citep{cPa16}.  Here, we point out a different  kind of difficulty. In certain regions of the parameter-configuration space  we show that determining whether an orbit will reenter due to the eccentricity growth mechanism or not is a difficult task due to the fractal boundaries of reentry basins. For such domains,  we explicitly quantify how uncertainties in the initial state must be reduced in order to achieve a corresponding reduction in prediction error, depending on the chosen mitigation scenario. \\

The remainder of the paper is structured as follows:
\begin{itemize}
	\item In Sec.\,\ref{sec:model}, we present the averaged (secular) Hamiltonian model we base our analysis on. The model includes Earth's $J_{2}$ perturbation and the SRP, both averaged in closed-form over the mean anomaly $M$ of the satellite. The secular model is prone to resonate. We also describe the numerical tools used to explore its associated dynamics. 
	\item In Sec.\,\ref{sec:chaoslifetime}, we numerically explore the system, focusing on two observables: the Smaller Alignment Index (SALI), a well-established chaos indicator, and the orbital lifetime computed over fine meshes of initial conditions. The parameter $A/m$ is allowed to take three distinct values in the set $\{1,5,10\}\,\mathrm{m}^2\mathrm{kg}^{-1}$, which, within the context of this study and relative to one another, are regarded as low, intermediate, and high area-to-mass ratios, respectively.  We present a cartography of regular and chaotic regions on a centennial timescale. We also find that the number of reentry orbits,  at intermediate and high $A/m$ values, follows an inverse-square power law in their lifetimes.  The long-term stability of such orbits is then reconsidered over a more practical and physically relevant timescale, which we call the effective timescale, prescribed by the reentry dynamics that obey the $25$-year guideline.    While the long-term dynamics display widespread chaos, the vast majority of reentry orbits show a regular behavior over the effective timescale. 
	\item In Sec.\,\ref{sec:boundaries}, we reveal that reentry basins may exhibit complex geometries depending on the targeted post-mission disposal guideline.  The complexity is assessed qualitatively and quantitatively by computing uncertainty exponents.   Our results reveal reentry regions with clear fractal structures, highlighting that the robust design of reentry trajectories requires dynamically informed criteria.
	\item Finally, Sec.\,\ref{sec:conclusion} summarizes our results, discusses  perspectives, and closes the paper.
\end{itemize}

\section{Secular Hamiltonian model and numerical tools}\label{sec:model}

In this section we present the secular Hamiltonian and the numerical tools used in our study.

\subsection{The secular Hamiltonian}

We consider a perturbed Hamiltonian formulation of the two-body problem, incorporating the perturbations due to the Earth's oblateness and SRP, as in \cite{iGk20}. In particular, we assume the so-called cannonball model, \ie that the Sun's radiation is normal to the surface of the space object at all times, and we neglect the effects of the Earth's albedo and shadowing. Since our focus is on the long-term evolution of the system, the Hamiltonian is averaged in closed form over the fast angle, namely the satellite's mean anomaly $M$ \citep{avKr97}. In terms of the Keplerian orbital elements $(a,e,i,\omega,\Omega,M)$, corresponding respectively to the semi-major axis, eccentricity, inclination, argument of perigee, longitude of the ascending node, and mean anomaly, the averaged Hamiltonian takes the form
\begin{align} 
	\label{eq:GkoliasHam-OE}
	{\mathcal{H}}
	= \frac{\mu r_E^2 J_2 \left( 1- 3\cos^2 i\right)}{4 a^3 \left(1 - e^2\right)^{3/2}} - \frac{3}{2} a e c_{R} P\frac{A}{m} \sum_{k=1}^6\mathcal{T}_k(i) \cos \psi_k(\omega,\Omega,\lambda_{\odot}).
\end{align} 
In Eq.\,\eqref{eq:GkoliasHam-OE}, $r_E$ denotes the Earth's radius and $J_2$ its oblateness coefficient. The parameter $c_R$ is the satellite's reflectivity coefficient, $A/m$ its area-to-mass ratio, and $P$ the solar radiation pressure constant at 1 astronomical unit. Furthermore, we set $\mu=\mathcal{G}m_E$, where $\mathcal{G}$ is the gravitational constant and $m_E$ the Earth's mass. The  $\mathcal{T}_k(i)$, for $k=1,\ldots,6$, are the coefficients of the six SRP-induced harmonics listed in Table\,\ref{tab:harmonics}. The obliquity of the ecliptic is fixed at $\varepsilon = 23.4\degree$, while the ecliptic longitude of the Sun evolves as $\lambda_{\odot} = \lambda_{\odot,0} + n_{\odot} t$, where $\lambda_{\odot,0}$ is determined from astronomical ephemerides and $n_{\odot}$ is the mean motion of the Sun at $2\pi/(86400\times365.25) \ \mathrm{s}^{-1}$. \\
\begin{table}
	\centering 
	\begin{tabular}{c|c|c}
		$k$ & $\mathcal{T}_k(i)$ & $\psi_k(\omega,\Omega,\lambda_{\odot})$ \\ \hline \hline
		1 & $\tfrac{1}{4}\,(\cos\varepsilon+1)(\cos i+1)$ & $ \omega+\Omega-\lambda_\odot$ \\ 
		2 & $-\tfrac{1}{4}\,(\cos\varepsilon+1)(\cos i-1)$ & $-\omega+\Omega-\lambda_\odot$ \\ 
		3 & $\tfrac{1}{2}\,\sin i \,\sin\varepsilon$ & $ \omega-\lambda_\odot$ \\ 
		4 & $-\tfrac{1}{2}\,\sin i \,\sin\varepsilon$ & $\omega+\lambda_\odot$ \\ 
		5 & $-\tfrac{1}{4}\,(\cos\varepsilon-1)(\cos i+1)$ & $\omega+\Omega+\lambda_\odot$ \\ 
		6 & $\tfrac{1}{4}\,(\cos\varepsilon-1)(\cos i-1)$ & $-\omega+\Omega+\lambda_\odot$ \\ \hline
	\end{tabular}
	\caption{The six harmonics governing the secular evolution of the Hamiltonian given in Eq.\,\eqref{eq:GkoliasHam-OE}.}
	\label{tab:harmonics}
\end{table}

For the Hamiltonian formulation, it is convenient to introduce the canonical Delaunay action-angle variables $(L,G,H,\ell,g,h)$, defined through 
\begin{equation}
	\left\{
	\begin{aligned}\label{eq:OEtoDel}
		L &= \sqrt{\mu a}, \quad & \ell &= M, \\
		G &= L \sqrt{1 - e^2}, \quad & g &= \omega, \\
		H &= G \cos i, \quad & h &= \Omega.
	\end{aligned}
	\right.
\end{equation}
The explicit time dependence introduced through the solar longitude $\lambda_{\odot}$ is removed by extending the phase space and introducing a dummy action $J$ conjugate to a variable $s$ such that $s(t)=n_{\odot}t+\lambda_{\odot,0}$.  The resulting autonomous Hamiltonian, describing a system with 3 degrees of freedom (DoF) in the variables $\vect{x} = (\vect{p},\vect{q})$ with $\vect{p} = (G,H,J)$ and $\vect{q} = (g,h,s)$, is finally written as
\begin{align}\label{eq:final_hamiltonian}
	\mathcal{H}(\vect{p},\vect{q})=
	h_{0}(\vect{p}) + h_{1}(\vect{p},\vect{q}),
\end{align}
with 
\begin{align}
	\label{eq:h0}
	h_{0}(\vect{p}) = 
	\frac{r_E^2J_2 \mu^4\left( G^2 - 3H^2 \right)}{4 G^5 L^3} + n_{\odot}J,
\end{align}
and
\begin{align}
	h_{1}(\vect{p},\vect{q})=
	- \frac{3 c_R P L^2 \frac{A}{m}}{2\mu} 
	\sqrt{1 - \frac{G^2}{L^2}} 
	\sum_{k=1}^{6} \mathcal{T}_k(i(L,G,H)) \cos \psi_k(g,h,s).
\end{align}
We note that since the angle $\ell$ is cyclic in Eq.\,\eqref{eq:final_hamiltonian}, the corresponding action $L$, and consequently the semi-major axis $a$, is conserved. Thus, the Hamiltonian depends parametrically upon $a$ and $A/m$.\\

The equations of motion associated with the Hamiltonian in Eq.\,\eqref{eq:final_hamiltonian} are expressed in canonical form as
\begin{align}
	\label{eq:eqmot}
	\dot{\vect{x}}=\mathcal{J}\nabla {\mathcal{H}}(\vect{x}), 
\end{align}
where the matrix $\mathcal{J}$ is the standard symplectic matrix $\mathcal{J} = 
\begin{pmatrix}
	0_3 & I_3 \\
	-\!I_3 & 0_3
\end{pmatrix}$ 
with $I_3$ and $0_3$ denoting the $3\times3$ identity and zero matrices, respectively. The gradient of the Hamiltonian is written as $\nabla {\mathcal{H}}(\vect{x}) = (\partial_{\vect{p}}{\mathcal{H}}, \partial_{\vect{q}}{\mathcal{H}})$.\\

Resonances strongly influence the dynamics associated with the equations of motion in Eq.\,(\ref{eq:eqmot}). For a wavenumber $\vect{k} \in \mathbb{Z}^{3}$ labeling a specific resonance, the resonance condition takes the form
\begin{align}\label{eq:res}
	\vect{k} \cdot 
	\varpi_{0}(\vect{p})=0, 
\end{align}
where 
$\varpi_{0}(\vect{p})=\partial_{\vect{p}}h_{0}(\vect{p})$ is the unperturbed frequency vector. This vector contains the frequency rates associated with $\dot{\omega}$, $\dot{\Omega}$ under the secular $J_{2}$ effect and the constant frequency $n_{\odot}$. The ``resonant manifolds'' associated to Eq.\,(\ref{eq:res}), defining a $2$D surface in the space of the actions $(L,G,H)$ -- equivalently, $(a,e,i)$ in orbital elements -- have been portrayed in several instances for low  values of $\vert \vect{k} \vert$, see \eg \cite{Alessi_MNRAS}.

\subsection{The Smaller Alignment Index method}

To distinguish between regular and chaotic motion in the 6-dimensional phase space of the Hamiltonian of Eq.\,\eqref{eq:final_hamiltonian}, we use the SALI method, first developed in \cite{chSk01}. This is a variational indicator based on the time evolution of the alignment of two deviation vectors from the studied orbit \citep{chSk03,chSk04,chSk16}, which has been successfully applied to studies of various dynamical systems \citep{aSz04,tMa11,jBo12,eeZ17,laCa18,eeZ21,yAl24,hMo24}. The evolution of a deviation vector $\vect{w}$ from a reference orbit is governed, at first order, by the variational equations
\begin{align}
	\label{eq:varEqns}
	\dot{\vect{w}}=\mathcal{J}\nabla^{2} \mathcal{H}(\vect{x}) \vect{w}. 
\end{align}

To quantify the chaoticity of an orbit with initial condition $\vect{x}_0$, we compute the SALI by simultaneously integrating the equations of motion (Eq.\,\eqref{eq:eqmot}) and the variational equations (Eq.\,\eqref{eq:varEqns}) for two initially linearly independent deviation vectors, $\vect{w}_0^{1}$ and $\vect{w}_0^{2}$. The SALI at time $t$ is defined as
\begin{align}
	\label{eq:SALI}
	\textrm{SALI}(\vect{x}_{0},\vect{w}_{0}^{1},\vect{w}_{0}^{2},t)
	=
	\min\big\{
	\norm{\hat{\vect{w}}^{1}(t)-\hat{\vect{w}}^{2}(t)},
	\norm{\hat{\vect{w}}^{1}(t)+\hat{\vect{w}}^{2}(t)}
	\big\},
\end{align}
where $\|\cdot\|$ denotes the Euclidean norm and the normalized deviation vectors are given by
\begin{align}
	\label{eq:SALI_norm_vect}
	\hat{\vect{w}}^{i}(t)=\frac{\vect{w}^{i}(t)}{\norm{\vect{w}^{i}(t)}}, \,\,\,i=1,2.
\end{align}
For brevity, we simply denote the index as $\textrm{SALI}(t)$. \\

In our study, the equations of motion  and the corresponding variational equations  are numerically integrated using the variable time-step integrator DOP853~\citep{eHa93}. The integrations are performed over timescales of up to $100$ years, corresponding to approximately $10^{5}$ orbital revolutions for the considered semi-major axis values. Throughout all simulations, the relative energy error typically remains below $10^{-8}$ ensuring the reliability of the numerical results. For the computation of the SALI, the initial deviation vectors are randomly drawn from a normal distribution and subsequently orthonormalized to unit length, so that $\textrm{SALI}(0)=\sqrt{2}$.\\

The distinction between regular and chaotic motion is efficiently captured by the SALI through its markedly different time evolution in the two cases. In particular, it has been shown in~\cite{chSk04} that, after a short transient, the SALI for chaotic orbits exhibits an exponential decay of the form
\begin{align}
	\label{eq:SALI_chaos}
	\textrm{SALI}(t) \propto e^{-(\lambda_{1}-\lambda_{2})t},
\end{align}
where $\lambda_{1} \geq \lambda_{2}$ are the orbit's two largest Lyapunov exponents. This behavior originates from the tendency of deviation vectors to align along the most unstable direction, associated with the maximal Lyapunov exponent. As a consequence, the SALI rapidly tends to zero, providing a clear signature of chaotic dynamics. In contrast, for regular motion in Hamiltonian systems with $N \geq 2$ DoF, deviation vectors eventually fall on the $N$-dimensional tangent space of the invariant torus. In this case, there is no mechanism enforcing their alignment along a single direction. As a result, the SALI does not decay to zero but instead fluctuates around a positive value, remaining practically constant on average~\citep{chSk03,chSk04}.\\

\begin{figure}
	\centering
	\includegraphics[width=0.6\linewidth]{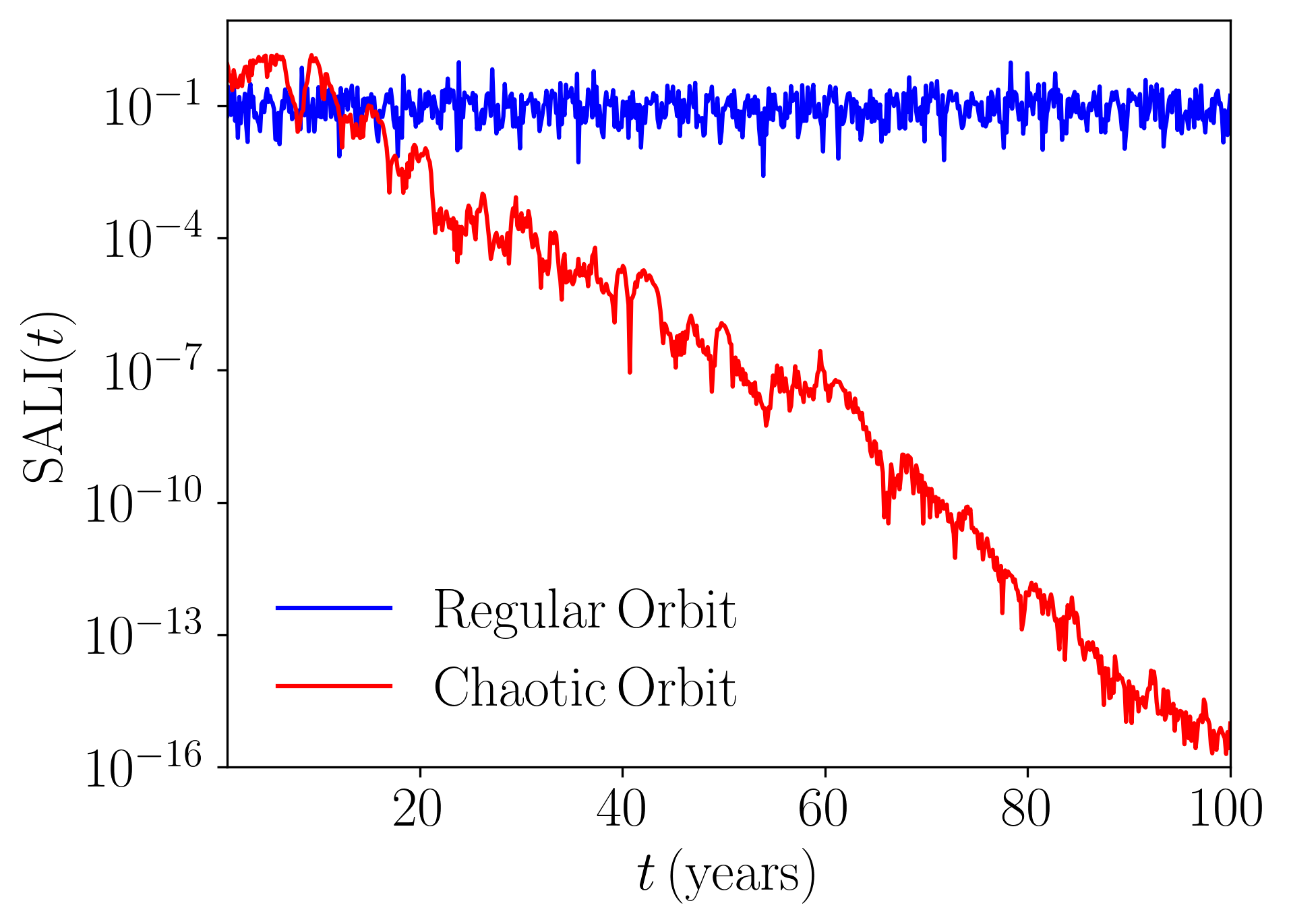}
	\caption{ 
		Time evolution of the SALI for a representative regular orbit (blue curve) with initial condition $(a_0,e_0,i_0, \omega_0,\Omega_0) = (11\,100\mathrm{km}, 0.05, 70\degree, 0\degree,0\degree)$ and a representative chaotic orbit (red curve) with initial condition $(a_0,e_0,i_0, \omega_0,\Omega_0) = (11\,100\mathrm{km}, 0.05, 100\degree, 0\degree,0\degree)$, for the Hamiltonian system of Eq.\,\eqref{eq:final_hamiltonian} with  $A/m=10\,\mathrm{m}^2\mathrm{kg}^{-1}$. The SALI of the chaotic orbit decays to machine precision exponentially fast while the regular orbit exhibits bounded fluctuations about some positive constant.}
	\label{fig:fig1}
\end{figure}

This qualitative difference (exponential decay to zero for chaotic orbits versus bounded, non-zero fluctuations for regular ones) renders the SALI a particularly efficient and reliable indicator for distinguishing between ordered and chaotic dynamics. This behavior is demonstrated in Fig.\,\ref{fig:fig1} showing the time evolution of the SALI for two representative orbits. The chaotic orbit (red curve) exhibits the characteristic rapid exponential decay of the index, which reaches machine precision over the timescale considered. In contrast, the regular orbit (blue curve) displays bounded fluctuations of the SALI around a nearly constant positive value. These two examples highlight the clearly different temporal behavior of the index in chaotic and regular regimes.

\section{Stability and orbital lifetimes}\label{sec:chaoslifetime}
In this section we discuss stability results in the long-term, the distribution of reentry orbits and the stability over an effective timescale.

\begin{figure}
	\centering
	\makebox[\textwidth][c]{%
		\includegraphics[width=1.3\textwidth]{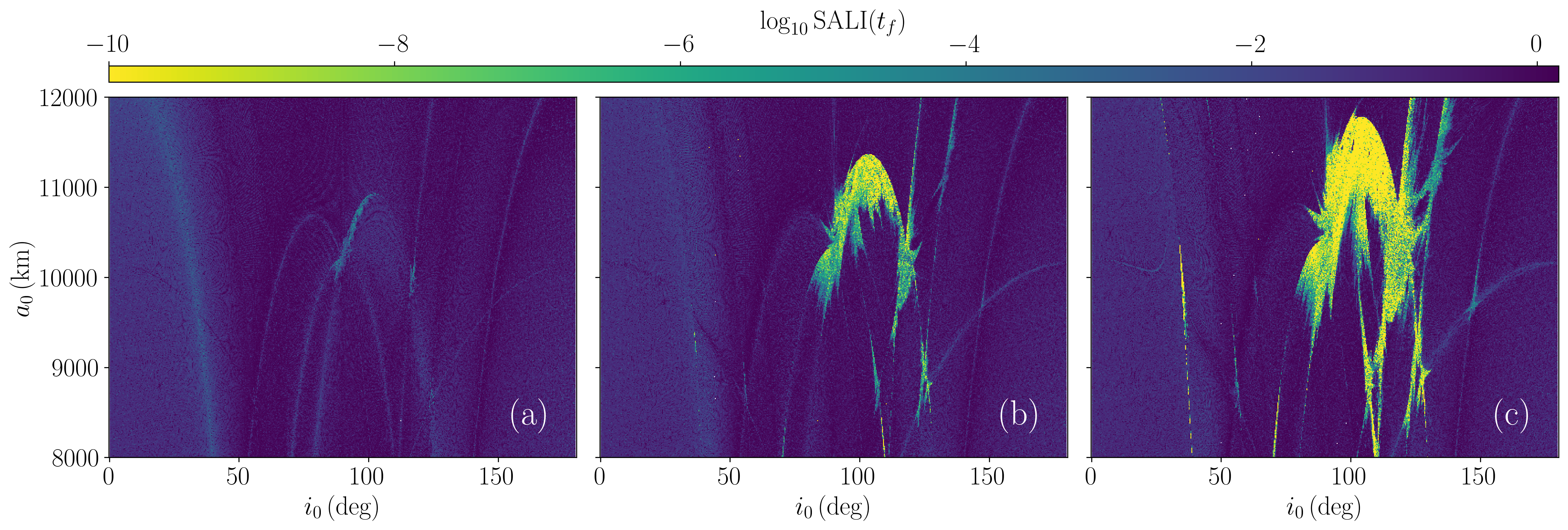}
	}
	\caption{
		Initial conditions in the $i-a$ space colored according to the $\log_{10}\mbox{SALI}$ value  computed at $t_f=100$ years, over the interval $a_0\in[8\,000\mathrm{km},12\,000\mathrm{km}]$ and $i_0\in(0\degree,180\degree)$, with $e_0=0.05$ fixed and $\omega_0=\Omega_0=\lambda_{\odot,0}=0\degree$, for (a)  $A/m=1\,\mathrm{m}^2\mathrm{kg}^{-1}$,  (b) $A/m=5\,\mathrm{m}^2\mathrm{kg}^{-1}$, and  (c) $A/m=10\,\mathrm{m}^2\mathrm{kg}^{-1}$. The presence of chaos increases with increasing $A/m$.
	}
	\label{fig:fig2}
\end{figure}

\subsection{Long-term stability}
\label{subsec:chaos}

Using the SALI indicator, we perform a stability analysis of the secular Hamiltonian in Eq.\,(\ref{eq:final_hamiltonian}) to reveal the geometry and extent of possible chaotic structures. To this end, taking into account that the phase space of the autonomous Hamiltonian has dimension 6, we compute dynamical maps over 2-dimensional slices of fine Cartesian meshes of initial conditions. From the computation of several dynamical maps parameterized by the angular variables, we noticed that the observed structures do not significantly change for different values of the system’s initial phases $\omega_0$, $\Omega_0$ and $\lambda_{\odot,0}$, as  maps presented in Fig.\,\ref{fig:appfig1} in Appendix\,\ref{app:A} denote. This allows us, in what follows, to present  results with  specific fixed values of initial angles, namely set to $\Omega_0 =0\degree$, $\omega_0 =0\degree$, and $\lambda_{\odot,0}=0\degree$. Thus, we have $3$ remaining coordinates, $(G,H,J)$, and $2$ free parameters ($a$ and $A/m$). The action $J$ has no actual dynamical meaning, leading to $\binom{4}{2}=6$ possible different coordinate planes. Here, we specifically focus on results in the $i-a$ space by freezing the initial eccentricity to $e_{0}=0.05$, underlining that the resonant manifolds of Eq.\,(\ref{eq:res}) do not strongly depend on the choice of $e_{0}$ for values $e_0\in\{0.001,0.005,0.01,0.05\}$ (see Fig\,\ref{fig:appfig2} in Appendix\,\ref{app:A}). The only parameter left, $A/m$, takes a value in the set $\{1,5,10\}\,\mathrm{m}^2\mathrm{kg}^{-1}$.\\

Fig.\,\ref{fig:fig2} shows three dynamical maps of the  region $i_0 \in (0\degree,180\degree)$ with an initial semi-major axis covering the range $a_0\in[8\,000\mathrm{km}, 12\,000\mathrm{km}]$ for increased values of the parameter $A/m$. For $A/m=1\,\mathrm{m}^2\mathrm{kg}^{-1}$ (panel (a)), the vast majority of orbits are regular  and do not show a chaotic signature on the considered timescale. We note, however, that SALI delineates several thin structures related to the resonances of Table\,\ref{tab:harmonics}. Increasing the value of  $A/m$ to $5\,\mathrm{m}^2\mathrm{kg}^{-1}$ (panel (b)), we observe the appearance of an inverted U-shaped, yellow-colored, region corresponding to connected chaotic seas, located around $i_0 \sim 100^{\circ}$ and $a_0 \sim 11\,000\mathrm{km}$.  Thinner chaotic structures emanate from the main sea along resonant lines. The chaotic structures grow further with the increase of $A/m$ to $10\,\mathrm{m}^2\mathrm{kg}^{-1}$ (panel (c)). Several chaotic filamentary structures emerge at higher and lower altitudes, predominantly along the resonant lines and across resonant junctions. In particular, highly-inclined orbits, with $i_0 \sim 110^{\circ}$, might now be chaotic at lower semi-major axis values, \eg at $a_0 \sim 8\,000\mathrm{km}$, and connect to the large chaotic sea at larger $a$ values. \\

For the sake of completeness, Fig.\,\ref{fig:appfig3} in Appendix\,\ref{app:B} and Fig.\,\ref{fig:appfig5} in Appendix\,\ref{app:C} complement the cartography by presenting dynamical maps in two other slices of the phase space, namely $i-e$ and $e-a$ space, respectively. Even in these spaces, we see that the presence of chaos grows with the increase of $A/m$.

\subsection{Orbital lifetimes}

Connections between chaotic transport and orbital lifetimes in Earth's orbits have been reported in many instances as the result of the  eccentricity growth mechanism, especially in medium Earth orbits, see \eg  \cite{aRo08,aRo16,iGk19}. In the following, an orbit is said to be a reentry orbit over the time window $[0,t_{f}]$ if it reaches it's critical eccentricity $e_{\star}$ at time $t_{\star} \leq t_{f}$, where $t_{f}$ is set to $100$ years. The critical eccentricity $e_{\star}$ is derived from the perigee altitude condition
\begin{align}
	r_{\pi}=a_{0}(1-e_{\star}) \ge r_{E} + \delta, 
\end{align} 
where $\delta$ is set to $120\mathrm{km}$. Solving for $e_{\star}$, one finds that
\begin{align}
	\label{eq:e_star}
	e_{\star} = 1 - \frac{r_{E}+\delta}{a_{0}}. 
\end{align}

In the following, we investigate the connection between chaos and orbital lifetime,  
restricting our analysis to reentry trajectories with orbital lifetimes below or equal to $25$ years, in accordance with standard mitigation guidelines \citep{IADC, NASA}.   
For this set of initial conditions, we further distinguish orbits achieving reentry within $5$ years, reflecting more stringent targets currently advocated \citep{FCC,ESA}.   \\

\begin{figure}
	\centering
	\includegraphics[width=1\linewidth]{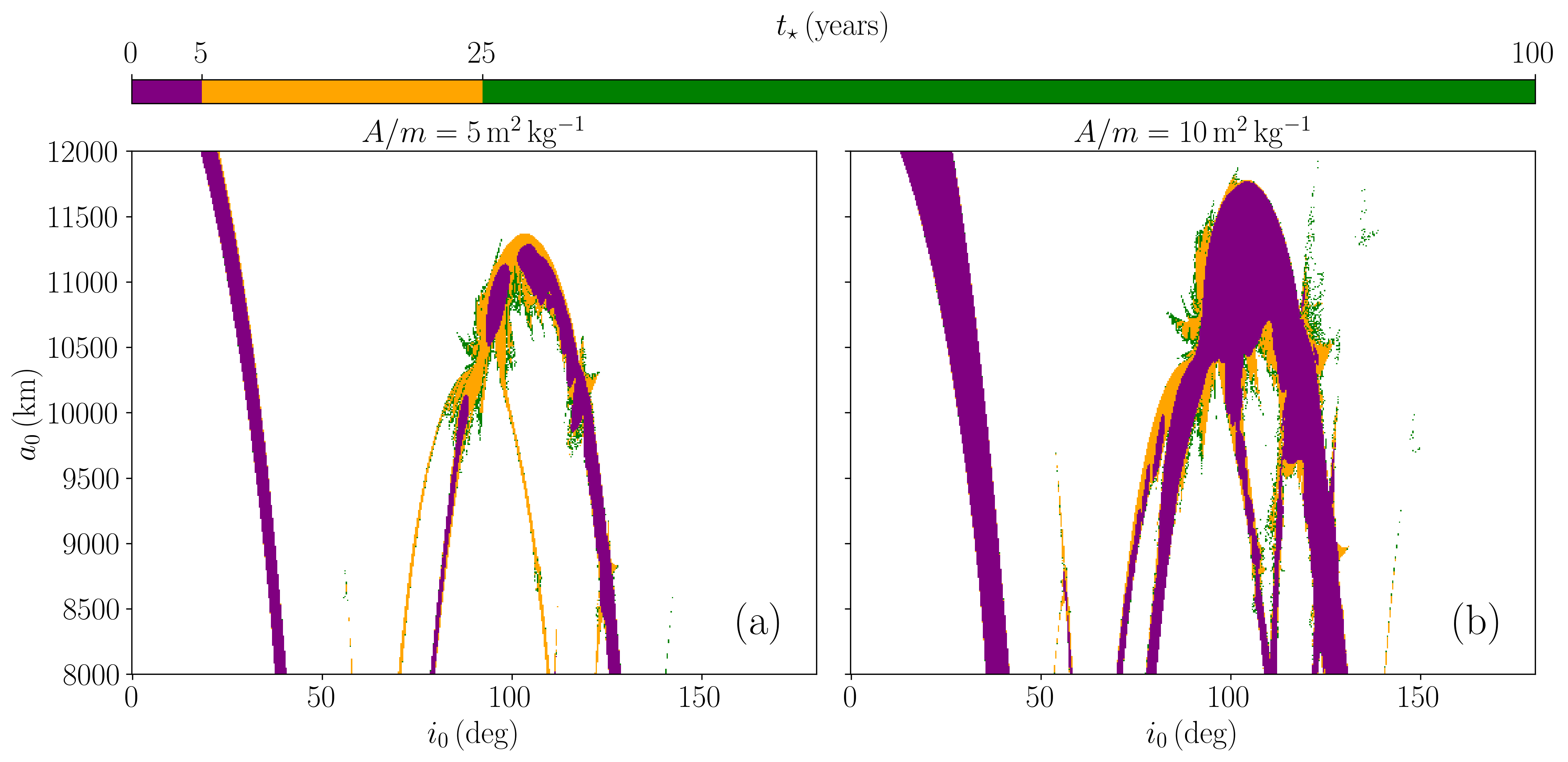}
	\caption{
		Identification of initial conditions in the $i-a$ space consistent with the $5$-year (violet) and $25$-year (orange) mitigation guidelines,
		for $e_{0}=0.05$ and $\omega_0=\Omega_0=\lambda_{\odot,0}=0\degree$, with (a) $A/m=5\,\mathrm{m}^2\mathrm{kg}^{-1}$ and (b) $A/m=10\,\mathrm{m}^2\mathrm{kg}^{-1}$. Initial conditions corresponding to orbits with reentry times $25 < t_\star \leq 100$ years are colored in green.
	}
	\label{fig:fig3}
\end{figure}

Fig.\,\ref{fig:fig3} shows the distribution of reentry orbits in the $i-a$ space, classified according to the two mitigation scenarios ($5$ and $25$ years), for  $A/m=5$ and  $10\,\mathrm{m}^2\mathrm{kg}^{-1}$. The case where $A/m=1\,\mathrm{m}^2\mathrm{kg}^{-1}$ is omitted due to the absence of significant structures, as not many orbits reenter within $25$ years for this value of $A/m$. We observe that the distribution of the reentry orbits in Fig.\,\ref{fig:fig3} is overall reminiscent of the chaotic regions depicted in Fig.\,\ref{fig:fig2}, particularly the inverted U-shaped structure at large $a_0$ and $i_0$ values.  Nevertheless, this correspondence is not one-to-one. The most noticeable example is the well delineated reentry corridor emanating from $i_0 \sim 40^{\circ}$ which corresponds to  $\dot{\psi}_1=0$. Interestingly enough, most of the reentry orbits comply with the stringent $5$-year rule. Orbits that fail to satisfy the $5$-year criterion but still comply with the $25$-year rule are predominantly found at large values of $a_0$, and tend to lie mostly near the boundaries of the reentry domain. Lifetime maps  shown in Fig.\,\ref{fig:fig3} for the $i-a$ space are also shown for the $i-e$ and $a-e$ spaces in Fig.\,\ref{fig:appfig4} and Fig.\,\ref{fig:appfig6} in Appendices \ref{app:B}  and \ref{app:C} respectively. \\

\begin{figure}
	\centering
	\includegraphics[width=1\linewidth]{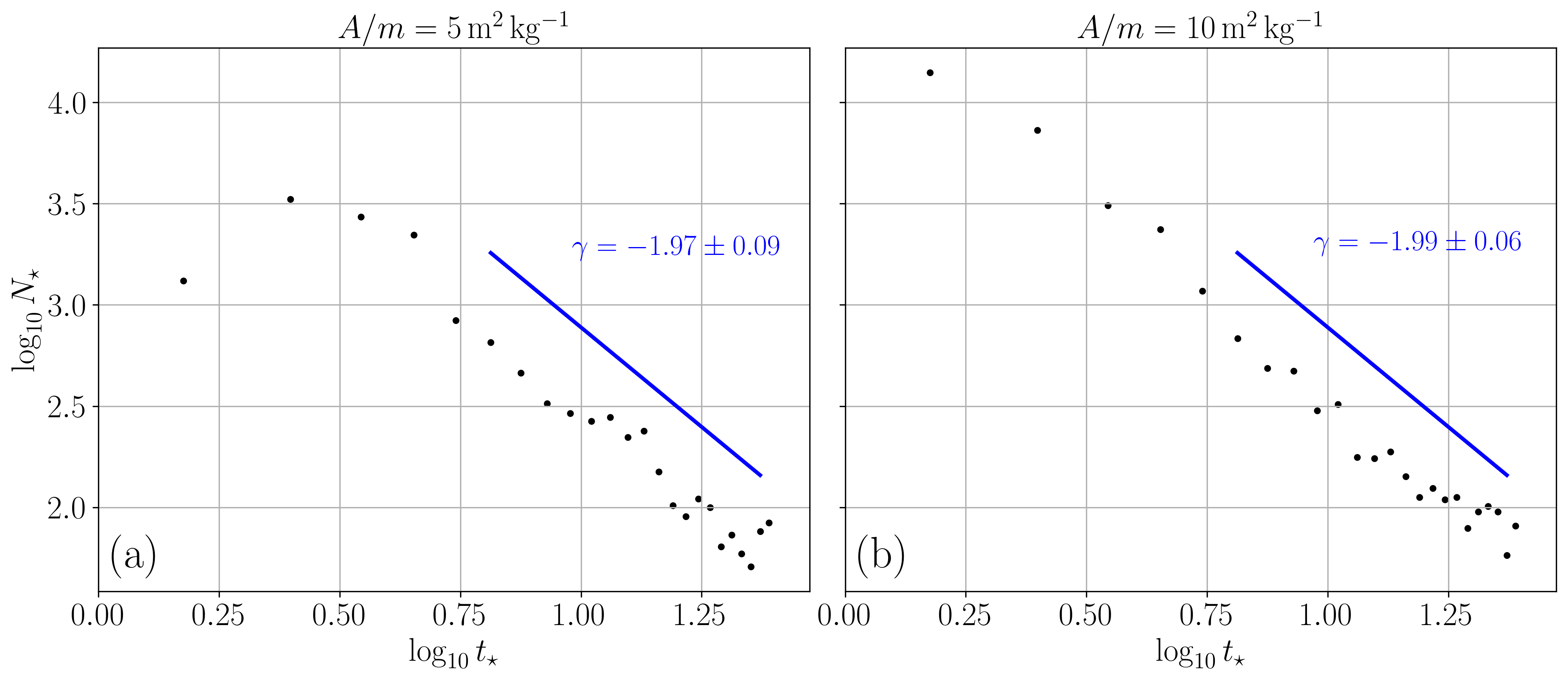}
	\caption{
		The number of  initial conditions $N_\star$ from Fig.\,\ref{fig:fig3} leading to orbits that reenter in the time window of $25$ years as a function of the reentry time $t_\star$, for (a) $A/m=5\,\mathrm{m}^2\mathrm{kg}^{-1}$ and (b) $A/m=10\,\mathrm{m}^2\mathrm{kg}^{-1}$. The results follow power laws with exponents $\gamma$ very close to $\gamma=-2$. 
	}
	\label{fig:fig4}
\end{figure}

We further characterize the reentry orbits by looking at the number $N_\star$ of reentry orbits at time $t_\star$. We find that $N_\star$ closely follows an inverse-square power law,
\begin{align}
	\label{eq:N_fit_t}
	N_\star \propto t_\star^{\gamma}, \quad \gamma \sim -2, 
\end{align} 
as shown in Fig.\,\ref{fig:fig4}. Both fits have a determination coefficient of $R^{2} > 0.96$ and small residuals. Table \ref{tab:powerlaws} extends this analysis to various $e_{0}$ values and reports  the corresponding exponents $\gamma$. This consistent value, $\gamma\sim-2$, provides a clear functional form that can be useful for predictions. 

\begin{table}
	\centering
	\begin{tabular}{c c c}
		\hline
		& $A/m = 5\,\mathrm{m}^2\mathrm{kg}^{-1}$ & $A/m = 10 \,\mathrm{m}^2\mathrm{kg}^{-1}$ \\
		$e_0$ & $\gamma$  & $\gamma$ \\
		\hline\hline
		0.001 & $-2.02\pm0.07$ & $-2.03\pm0.06$\\
		0.005 & $-1.98\pm0.07$ & $-2.04\pm0.05$\\
		0.01  & $-1.99\pm0.07$ & $-2.05\pm0.05$\\
		0.05  & $-1.97\pm0.09$ & $-1.99\pm0.06$ \\
		\hline
	\end{tabular}
	\caption{
		Power law exponents $\gamma$ (Eq.\,\eqref{eq:N_fit_t}) as a function of $e_{0}$.
		All numerical fittings have a determination coefficient of $R^2>0.96$.}
	\label{tab:powerlaws}
\end{table}

\subsection{Effective stability}

In subsection \ref{subsec:chaos}, regularity and chaoticity  were assessed at a rather large time set to $100$ years. In this section, we instead use the SALI to characterize the nature of an orbit at a  more practically relevant time $\tau$, referred to as the \textit{effective time}. Over the set of reentry trajectories that comply with the $25$-year rule, $\mathcal{R}_{25}$, where
\begin{align}
	\mathcal{R}_{25}
	=
	\{
	\vect{x}_{0} \, \vert \, t_{\star}(\vect{x}_{0}) \le 25\,[\mathrm{years}]
	\},
\end{align}
we define the effective time $\tau$ of an initial condition $\vect{x}_{0}$ as
\begin{align}\label{eq:define_tau}
	\tau: \mathcal{R}_{25}  &\to (0,25] \\
	\vect{x}_{0}    & \mapsto \notag \tau(\vect{x}_{0})=t_{\star}. 
\end{align}
Thus, the effective time has an upper bound of $25$ years over the timescale of interest. Moreover, if a reentry occurs earlier, the orbital behavior is characterized over the reentry timescale. By recording the value of the SALI at time $\tau$, we can classify the orbits as regular or chaotic at the more practically effective time governed by the reentry dynamics.  We underline that this approach is facilitated by the intrinsic design of the SALI, which monitors the instantaneous alignment of deviation vectors, reflecting in this way the local nature of the orbit, contrarily to, for example, the maximal Lyapunov exponent which averages the growth of tangent vectors over time.   \\

\begin{figure}
	\centering
	\includegraphics[width=1\linewidth]{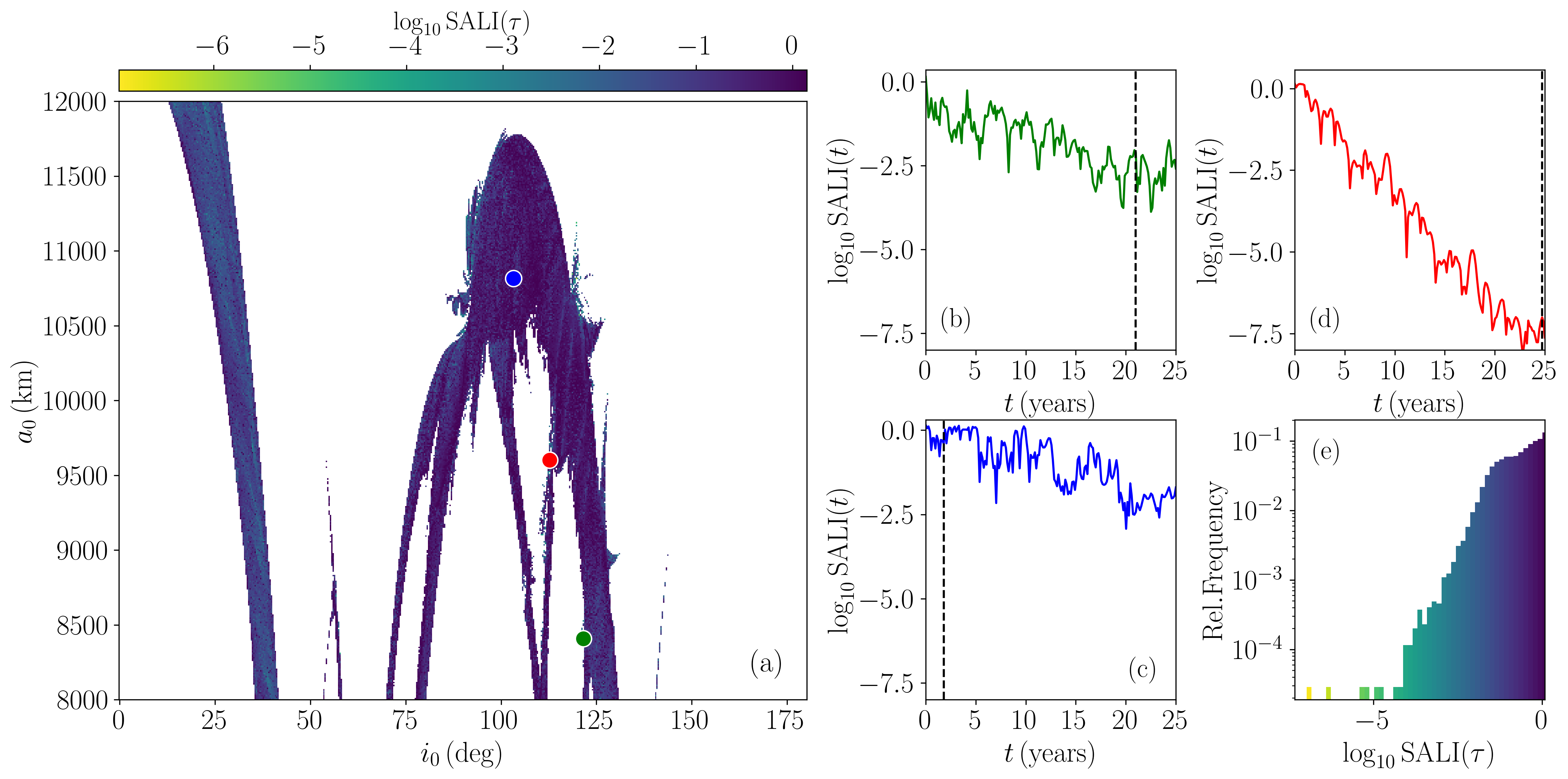}
	\caption{
		(a) Stability of the reentry orbits of Fig.\,\ref{fig:fig3}(b) at the effective time $\tau$ (Eq.\,\eqref{eq:define_tau}) for $A/m=10\,\mathrm{m}^2\mathrm{kg}^{-1}$. Over the reentry process, almost all orbits are regular, as the coloring of initial conditions according to their $\log_{10}\mbox{SALI}(\tau)$ indicates. Specific initial conditions are denoted by green, blue and red dots. The time evolution of SALI$(t)$ of these three orbits are respectively shown in panels (b), (c) and (d). Reentry times are marked by the dashed vertical lines in these panels. The orbits considered in (b) and (c) behave regularly until their reentry time with the one in (b) requiring more time to reenter. The orbit considered in (d) is a chaotic orbit which  reenters at $24.7$ years. In (e) we see the distribution of the $\log_{10}\mbox{SALI}(\tau)$ values for the orbits displayed in (a).
	}
	\label{fig:fig5}
\end{figure}

The composite plot of Fig.\,\ref{fig:fig5} shows that most of the orbits previously identified in Fig.\,\ref{fig:fig3}(b) are regular over the effective time interval. More specifically, among all reentry orbits, 99.7\% for $A/m=5\,\mathrm{m}^2\mathrm{kg}^{-1}$ and 99.8\% for $A/m=10\,\mathrm{m}^2\mathrm{kg}^{-1}$ have $\log_{10}(\rm{SALI}(\tau)) \ge -3$. In addition to the SALI dynamical map evaluated at the effective time (panel (a)), the time evolutions of the SALI of some representative orbits (panels (b), (c) and (d)) are also shown. In panels (b), (c) and (d) the reentry time is denoted by a dashed vertical line. Panels (b) and (c) display the prototypical behavior of the index for regular motion (see Fig.\,\ref{fig:fig1}). Although the orbit in panel (b) has a longer reentry time than that of panel (c), both orbits are regular up to their respective reentry times. In contrast, the orbit in panel (d) exhibits chaotic behavior at the time of reentry (approximately, $24.7$ years). This orbit is one of the very few effectively chaotic orbits present in panel (a), as further confirmed by panel (e), which shows the distribution of $\log_{10}\mathrm{SALI}(\tau)$ values in colors corresponding to panel (a). The scarcity of values below $\log_{10}\mathrm{SALI}(\tau)=-5$ further supports the predominance of regular behavior over  the effective timescale. Thus, the results of Fig.\,\ref{fig:fig5} clearly demonstrate that over the effective timescale, the reentry dynamics of almost all orbits are regular. \\

\begin{figure}
	\centering
	\includegraphics[width=1\linewidth]{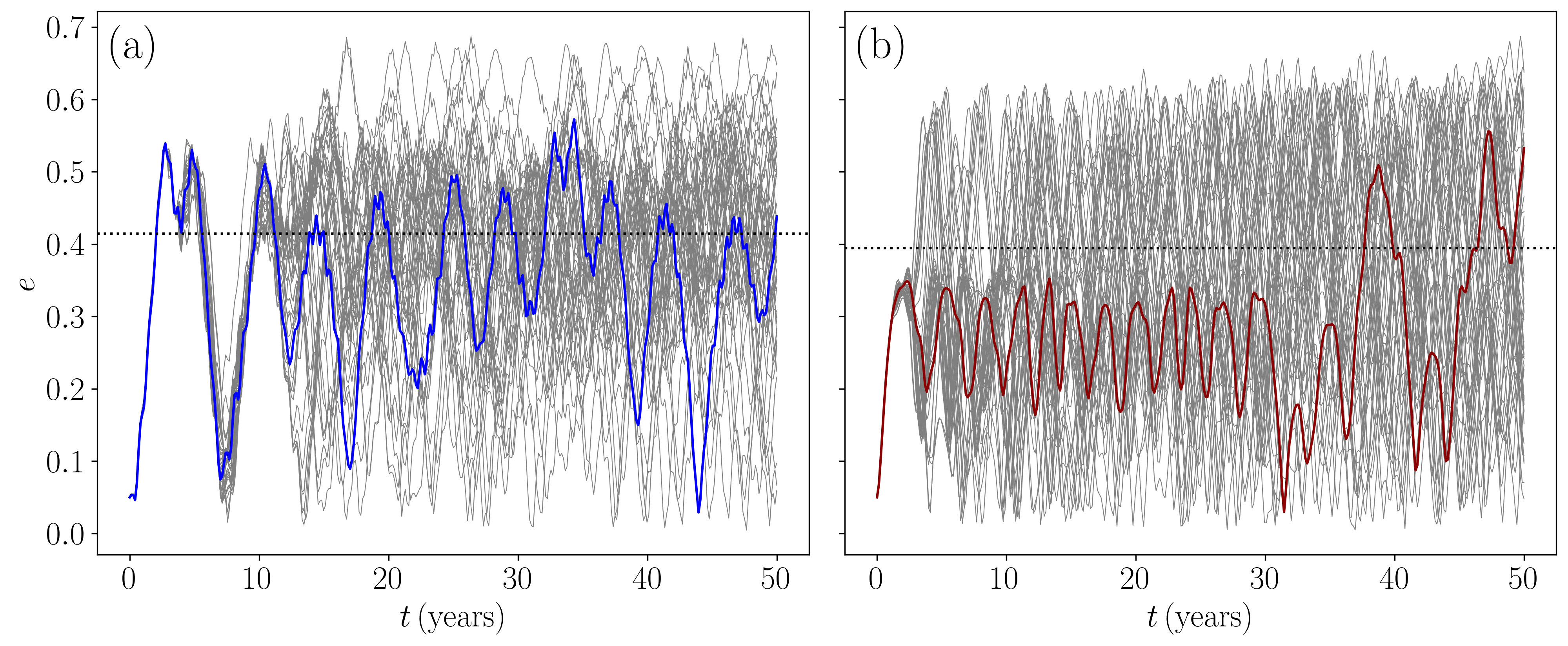}
	\caption{Evolution of the eccentricity of $50$ nearby randomly generated initial conditions (shown in gray) in a neighborhood of a nominal orbit (blue or red). The nominal orbit in (a) corresponds to the initial condition $(a_0,e_0,i_0,\omega_0,\Omega_0)=(11\,100\mathrm{km}, 0.05, 100\degree, 0\degree, 0\degree)$ (blue) and in (b) to $(a_0,e_0,i_0,\omega_0,\Omega_0)=(10\,727.98\mathrm{km}, 0.05, 118.36\degree, 0\degree, 0\degree)$ (red). The dotted horizontal line corresponds to the value of the respective nominal orbit's critical eccentricity $e_\star$ (Eq.\,\eqref{eq:e_star}). All orbits in (a) and (b) have $A/m=10\,\mathrm{m}^2\mathrm{kg}^{-1}$.}
	
	\label{fig:fig6}
\end{figure}

To further illustrate this point, Fig.\,\ref{fig:fig6} shows the long-term evolution (up to $50$ years) of the eccentricity of two nominal chaotic orbits. In gray, the evolutions of a cluster of 50 orbits near to the respective nominal orbits are shown. These neighboring initial conditions are drawn from a two-dimensional normal distribution and further constrained to lie within $1\degree$ and $1\,\mathrm{km}$ of the corresponding nominal orbit. Despite the chaotic nature of both nominal orbits, we observe a clear divergence of the nearby orbits from the nominal one, occurring at approximately 10 years for the blue orbit in panel (a) and around 3 years for the red orbit in panel (b). Nevertheless, in the case of panel (a), the nominal orbit remains effectively regular up to its reentry time, with no appreciable divergence of nearby orbits prior to this event. Importantly, this evolution complies with the 5-year rule. For this case, the separation of nearby orbits becomes evident only after the crossing of the critical eccentricity (indicated by dotted horizontal lines in both panels of Fig.\,\ref{fig:fig6}). This is an example of an eventually chaotic orbit which behaves regularly on the effective timescale. In contrast, the nominal orbit in panel (b) exhibits a much earlier divergence of neighboring orbits, revealing its chaotic character well before reentry. We note that this orbit does not satisfy the 25-year mitigation requirement and therefore does not appear in Fig.\,\ref{fig:fig5}, as it reaches the critical eccentricity at approximately $40$ years.

\section{Smooth and fractal boundaries of  reentry basins}
\label{sec:boundaries}

While the previous sections have focused on the regularity of transport and the identification of ordered and chaotic regions, the present section explores the geometry of reentry basins. \\

\begin{figure}
	\centering
	\includegraphics[width=0.7\linewidth]{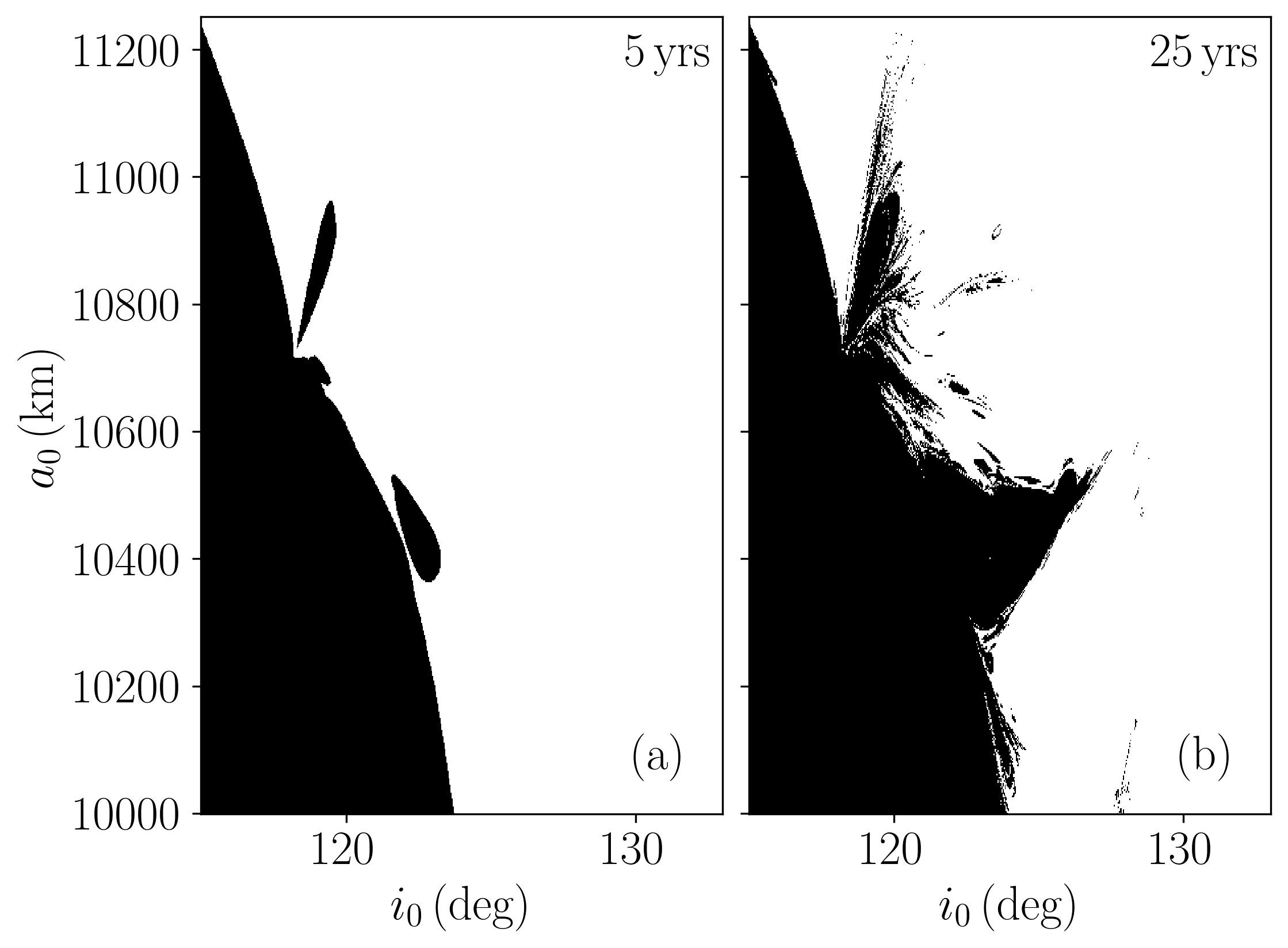}
	\caption{
		Distribution of  initial conditions of reentry orbits (black points) complying with the (a) $5$-year and (b) $25$-year mitigation guideline, over $a_0\in[10\,000\mathrm{km}, 11\,250\mathrm{km}]$ and $i_0\in[115\degree, 133\degree]$ with $e_0=0.05$ and $\omega_0=\Omega_0=\lambda_{\odot,0}=0\degree$, for $A/m=10\,\mathrm{m}^2\mathrm{kg}^{-1}$. The reentry basin boundaries exhibit increasing structural complexity depending on the mitigation scenario. A visual inspection of the boundaries suggests fractality, something which is confirmed and quantified through the computation of uncertainty exponents.
	}
	\label{fig:fig7}
\end{figure}

The boundaries of reentry basins develop in a complex way, depending on the mitigation guideline chosen. An example is provided in Fig.\,\ref{fig:fig7} for the $5$-year and  $25$-year scenarios, where initial conditions leading to reentry orbits are colored in black. In the more restrictive 5-year case, the basin boundaries appear relatively smooth.  In contrast, for the $25$-year scenario, the boundaries become considerably more structured and complex, showing multiple intricate patterns on small scales. This property indicates that even tiny variations in the initial conditions might influence whether an orbit leads to reentry or not. To quantify and characterize this complex transition witnessed over several domains, we employ the uncertainty exponent method, often traced back to \cite{cGr83}, and  subsequently applied in  various  dynamical systems contexts \citep{mMu18,acMa19,mrSa22}. As our results will demonstrate, distinct fractal-like structures emerge in the basin boundaries depending on the mitigation scenario. In the following, we continue our focus on domains  within the $i-a$ space, considering the case  $A/m=10\,\mathrm{m}^2\mathrm{kg}^{-1}$, with $e_0=0.05$ and $\omega_0=\Omega_0=\lambda_{\odot,0}=0$ as before. \\

The uncertainty method compares labels assigned by a given observable to two nearby initial conditions, namely $\vect{x}_{0} \pm \bm{\epsilon}$, where $\bm{\epsilon}$ represents a small perturbation of the considered initial condition $\vect{x}_{0}$. In our case, the observable assigns a binary label to each initial condition depending on whether the corresponding orbit satisfies the prescribed mitigation guideline. The initial condition $\vect{x}_{0}$	is said to be \textit{$\bm{\epsilon}$-uncertain} if the two perturbed initial conditions $\vect{x}_{0} + \bm{\epsilon}$ and $\vect{x}_{0}-\bm{\epsilon}$ receive different labels under the action of this observable. \\

\begin{figure}
	\includegraphics[width=1\linewidth]{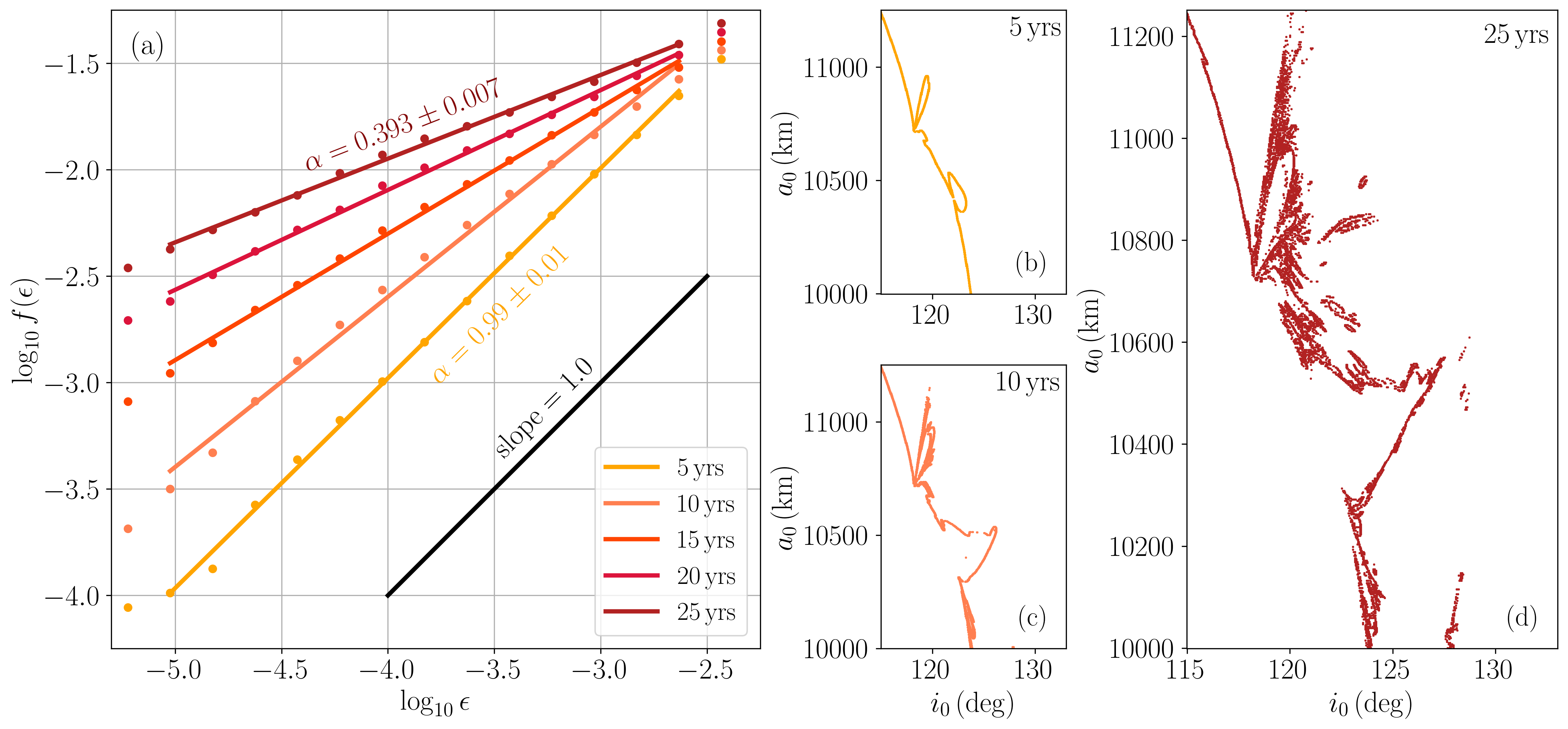}  
	\caption{\label{fig:fig8}
		(a) Slopes of the linear fits reveal the uncertainty exponents $\alpha$ (Eq.\,\eqref{eq:alpha}) computed for a series of mitigation guidelines for domain $\mathcal{D}$ (see Fig.\,\ref{fig:fig7} and Table \ref{tab:alpha}). Fractality becomes progressively more pronounced for less restrictive mitigation scenarios, with an uncertainty exponent close to $\alpha = 1$ for the $5$-year case, decreasing to $\alpha = 0.393$ for the $25$-year scenario, confirming the emergence of fractal basin boundaries. The corresponding transition is qualitatively reflected in the structure of the set of uncertain points presented in (b), (c), and (d) for the 5-, 10- and 25-year scenarios respectively, which range from smooth to increasingly complex patterns. The uncertain points shown in (b), (c) and (d) are computed using $\epsilon \approx 0.034\degree$.
	} 
\end{figure}

Over discretized domains $\mathcal{D}$ of initial conditions in the $i-a$ space containing substantial reentry boundaries, we computed the uncertainty fraction 
\begin{align}
	f(\bm{\epsilon})= \frac{N_{\textrm{uncertain}}(\bm{\epsilon})}{N},
\end{align} 
over a range of $\bm{\epsilon}$ values, setting $A/m=10\,\mathrm{m}^2\mathrm{kg}^{-1}$ throughout. In this investigation, we  perturb the initial condition only in the $i$-direction, so that $\bm{\epsilon}$ has a nonzero component only in $i$; we therefore denote it simply by $\epsilon$. The values of $\epsilon$ used to generate the best fit lines in Fig.\,\ref{fig:fig8} consist of 13 logarithmically spaced points between $(5.4\times 10^{-4})\degree$ and $0.13\degree$. The quantities $N_{\textrm{uncertain}}(\epsilon)$ and $N$ denote, respectively, the number of $\epsilon$-uncertain initial conditions  and the total number of sampled initial conditions in $\mathcal{D}$. As established in \cite{cGr83}, the uncertainty fraction $f(\epsilon)$ scales as 
\begin{align}
	\label{eq:alpha}
	f(\epsilon) \propto \epsilon^{\alpha},
\end{align}
where  $\alpha$ is called the uncertainty exponent. For smooth boundaries $\alpha=1$, whilst $0<\alpha<1$ indicates fractal basin boundaries. \\

This power law scaling has important practical implications for  reentry predictions when $\alpha \neq 1$. For illustrative purposes, let us consider the case $\alpha = 1/2$. A reduction of the initial condition uncertainty by a factor $100$ then leads to a reduction in the fraction of uncertain points by a factor of only $10$. In other words, substantial improvements in the precision of the initial conditions translate into limited gains in the reliability of the predicted label. This situation worsens for smaller $\alpha$, \ie for increasingly fractal boundaries.  \\

The composite plot  in Fig.\,\ref{fig:fig8} illustrates how the uncertain regions (\ie the sets of uncertain points) evolve as the mitigation guideline becomes progressively less restrictive. For pedagogical purposes, we consider scenarios corresponding to $5$-, $10$-, $15$-, $20$- and $25$-year thresholds. Comparing Fig.\,\ref{fig:fig7} and Fig.\,\ref{fig:fig8}, we observe a strong correlation between the  uncertain points computed from the boundary of the reentry set and the points lying on the boundary itself, at least for the cases shown in panels (b), (c) and (d) of Fig.\,\ref{fig:fig8}, which correspond to $\epsilon\approx 0.034\degree$. From the results of Fig.\,\ref{fig:fig8} it is evident that, in the $5$-year case the set of uncertain points forms a smooth curve. In contrast, in the $10$-year, and especially the $25$-year scenario, the set of uncertain points develops a significantly more intricate structure. This qualitative observation  is confirmed quantitatively through the estimation of the associated uncertainty exponents $\alpha$. The values of $\alpha$ reveal  a sharp transition in the power laws, ranging from smooth basin boundaries (with $\alpha$ close to $1$) to strongly fractal ones (reaching $\alpha=0.393$) as seen in panel (a) of Fig.\,\ref{fig:fig8}. We note that the linear fits in that panel have $R^{2}$ coefficients ranging from $0.993$ to $0.999$. Overall, in the considered $i-a$ domain, the variation in uncertainty exponents demonstrates that compliance with the $25$-year reentry guideline requires significantly higher precision in the initial conditions than satisfying the $5$-year criterion. While this result is consistent with intuition, the present analysis provides a quantitative and rigorous characterization of this dependence.\\

\begin{figure}
	\includegraphics[width=1\linewidth]{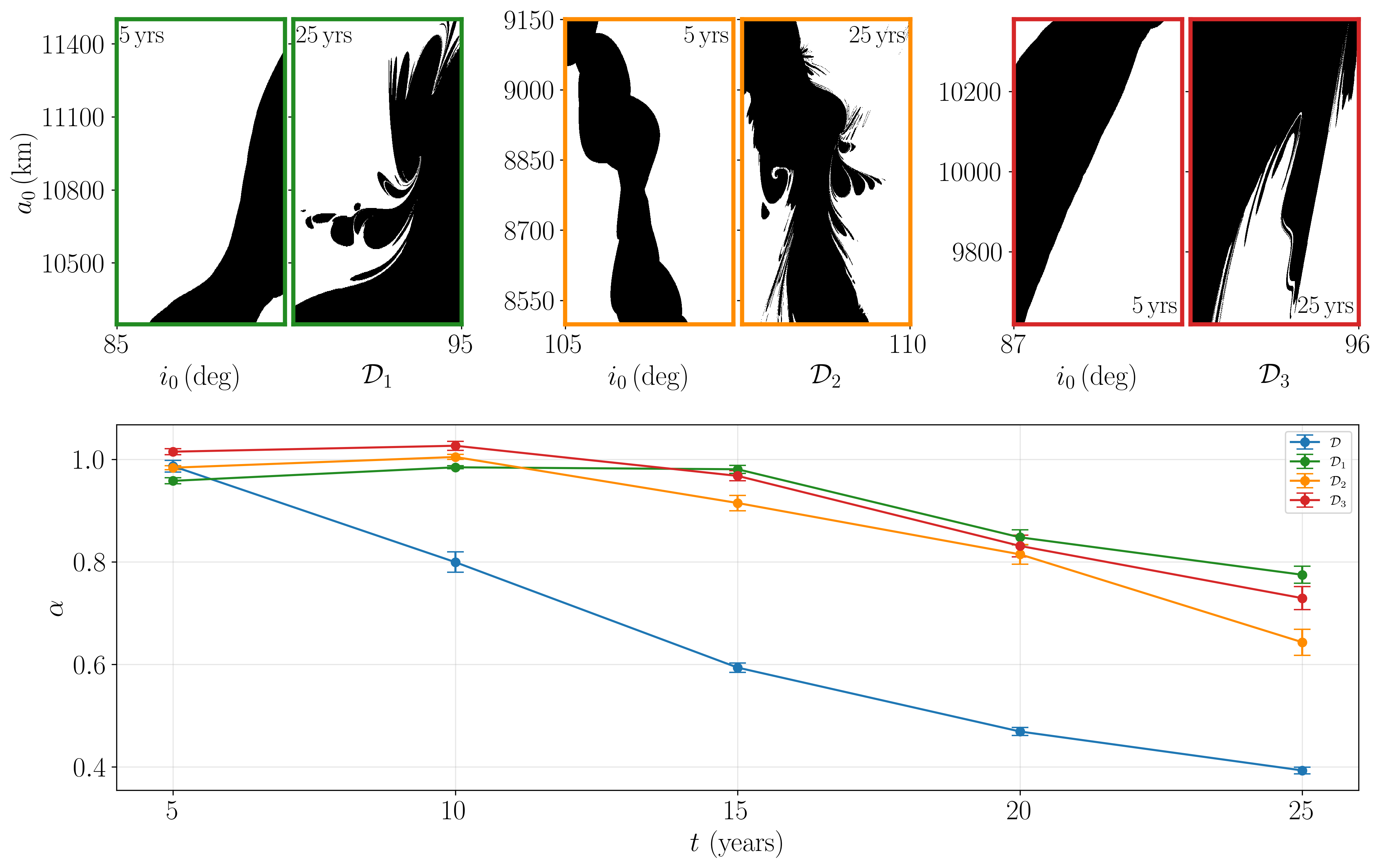} 
	\caption{\label{fig:fig9}
		Upper row: Reentry basins denoted by black points for the domains $\{\mathcal{D}_{i}\}_{i=1}^{3}$ of Table \ref{tab:alpha}. Lower panel: Uncertainty exponent $\alpha$ (Eq.\,\eqref{eq:alpha}) as a function of the years considered in the various mitigation scenarios, for the domains reported in Table \ref{tab:alpha}. The basins exhibit a clear fractality for the $25$-year mitigation scenario.
	} 
\end{figure}

We repeated the same computation and analysis over several domains $\mathcal{D}_{1}$, $\mathcal{D}_{2}$ and $\mathcal{D}_{3}$, and obtained qualitatively and quantitatively similar results. Although the values of $\alpha$ vary from one domain to another, the presence of fractal basin boundaries remains clearly pronounced, with exponents $\alpha <1$ in all cases for the $25$-year scenario. The definitions of the considered domains $\{\mathcal{D}_{i} \}_{i=1}^{3}$, together with the corresponding results,  are reported in Table \ref{tab:alpha}. The  evolution of the reentry sets for the extreme $5$- and $25$-year scenarios, along with their uncertainty exponents  as a function of time, are shown in Fig.\,\ref{fig:fig9}.\\

\begin{figure}
	\includegraphics[width=1\linewidth]{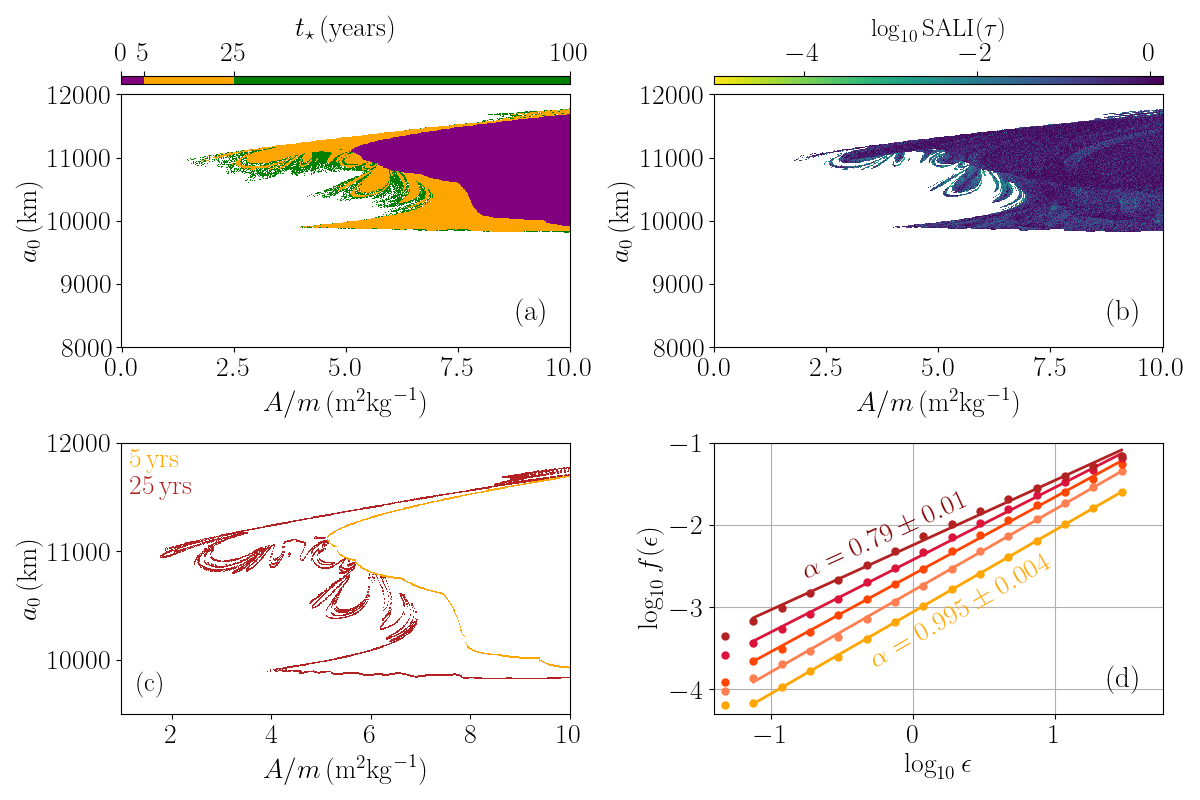}  
	\caption{\label{fig:fig10}
		Panel (a) identifies initial conditions complying with the 5-year (violet), 25-year (orange) mitigation guidelines, for initial conditions in the parameter space $A/m-a$  where $A/m\in[0.001\mathrm{m}^2\mathrm{kg}^{-1}, 10\mathrm{m}^2\mathrm{kg}^{-1}]$ and $a_0\in[8\,000\mathrm{km},12\,000\mathrm{km}]$, for $e_0=0.05$, $i_0=100\degree$, and $\omega_0=\Omega_0=\lambda_{\odot,0}=0\degree$. Initial conditions corresponding to orbits with reentry times $25 < t_\star \leq 100$ years are colored in green. Panel (b) shows the stability of the reentry orbits at the effective time $\tau$ (Eq.\eqref{eq:define_tau}) for the same initial conditions as in (a). Once again, evidenced by the predominance of high $\log_{10}\mathrm{SALI}(\tau)$ values, all orbits are regular at least up until reentry. Panel (c) maps the uncertain points calculated from a  perturbation in $a_0$ at $\epsilon \approx 4.68\mathrm{km}$ for the 5- and 25-year mitigation guidelines, showing the transition from smooth to fractal-like boundaries qualitatively. Panel (d) reports the slopes of the linear fits which reveal the uncertainty exponents $\alpha$ (Eq.\eqref{eq:alpha}) of the boundaries in the parameter space, for the same domain as in panel (c), with colors as those in Fig.\,\ref{fig:fig8}, for the boundaries corresponding to the 5-, 10-, 15-, 20-, 25-year mitigation guidelines. For less strict mitigation guidelines, fractality is more pronounced, as indicated by the $\alpha<1$ value  for the 25-year guideline and  $\alpha \approx 1$ for the 5-year guideline. 
	} 
\end{figure}

To further assess the robustness of our results, we extended the analysis to the parameter space $A/m-a$ for frozen metrical and angular variables. The view offers a substantially finer sampling in $A/m$, compared with the three values $\{1,5,10\} \mathrm{m}^2\mathrm{kg}^{-1}$ coarsely considered so far, providing a more detailed view of the evolution of the uncertain sets across the parameter space. The transition from smooth to increasingly intricate structures of reentry basins, reflected in the uncertainty exponent, is clearly not a consequence of the chosen $A/m$ used previously. The results of Fig.\,\ref{fig:fig10} therefore provide an additional confirmation of the properties of the reentry basins identified above.

\begin{table}
	\centering
	\begin{tabular}{c|cccc}
		& $\alpha$ of $\mathcal{D}$ & $\alpha$ of $\mathcal{D}_{1}$ & $\alpha$ of $\mathcal{D}_{2}$
		& $\alpha$ of $\mathcal{D}_{3}$
		\\
		\hline 
		\hline
		$5$ years & $0.99 \pm 0.01$ & $ 0.958\pm 0.006$ & $ 0.984\pm 0.004$  & $1.015 \pm 0.006$\\
		$10$ years & $0.8 \pm 0.02$ & $0.986 \pm 0.002$ & $1.005 \pm 0.005$ & $1.027\pm 0.009$\\
		$15$ years & $0.594 \pm 0.009$ & $0.981 \pm 0.007$ & $0.92 \pm 0.01$ & $0.97\pm 0.01$\\
		$20$ years & $0.469 \pm 0.008$ & $0.85 \pm 0.01$ & $0.81 \pm 0.02$ & $0.83 \pm 0.02$ \\
		$25$ years & $0.393 \pm 0.007$ & $0.77 \pm 0.02$ & $0.64 \pm 0.03$ & $0.73 \pm 0.02$ \\
		\hline \hline 
	\end{tabular}
	\caption{\label{tab:alpha}
		Uncertainty exponents $\alpha$ (Eq.\,\eqref{eq:alpha}) computed over various domains  in the $i-a$ space according to various mitigation guidelines. We namely consider the domains $\mathcal{D}=[115\degree,133\degree] \times [10\,000\mathrm{km},11\,250\mathrm{km}]$ (Fig.\,\ref{fig:fig7}), together with $\mathcal{D}_{1}=[85\degree,95\degree] \times [10\,250\mathrm{km},11\,500\mathrm{km}]$, $\mathcal{D}_{2}=[100\degree, 110\degree] \times [8\,500\mathrm{km},10\,000\mathrm{km}]$, and $\mathcal{D}_{3}=[87\degree, 96\degree] \times [9\,620\mathrm{km}, 10\,380\mathrm{km}]$ (Fig.\,\ref{fig:fig9}), all having $e_0=0.05$, $\omega_0=\Omega_0=\lambda_{\odot,0}=0\degree$ and $A/m=10\,\mathrm{m}^2\mathrm{kg}^{-1}$. In general, for mitigation timescales above  10 years the uncertainty exponents are significantly smaller than $1$, indicating the fractal nature of the reentry basin boundaries.}
\end{table}

\section{Conclusions}
\label{sec:conclusion}

Based on an averaged model incorporating Earth’s $J_{2}$ oblateness and SRP, the present study complemented and supplemented several aspects of the problem. The main contributions and conclusions are summarized as follows:
\begin{enumerate}
	\item The secular Hamiltonian has $3$ DoF and depends parametrically on two quantities: the secularly invariant semi-major axis $a$ and the area-to-mass ratio $A/m$. To visualize the onset and extent of chaos within this high-dimensional space, the SALI chaos indicator was used, focusing primarily on sections in the $i-a$ space (with results in the $e-i$ and $a-e$ spaces reported in Appendix\,\ref{app:B} and \ref{app:C}, respectively) for several $A/m$ values. We showed numerically that the extent of chaotic motion increases with increasing $A/m$, while the overall dynamical structures remained practically unchanged for different initial phases of the system (see Appendix\,\ref{app:A}). 
	\item For $A/m \ge 5\,\mathrm{m}^2\mathrm{kg}^{-1}$, the number of orbits reaching their critical eccentricity within $25$ years follows an inverse-square power law with respect to the corresponding reentry time. Moreover, for sufficiently large values of $A/m$, most reentry orbits reach their critical eccentricity within only a few years.
	\item Despite the potential for long-term chaotic behavior, the  majority of orbits that reach their critical eccentricity within $25$ years exhibit regular dynamics until their reentry.
	\item By varying the mitigation guideline from $5$ to $25$ years, we found that the associated basins of reentry orbits develop increasingly pronounced fractal-like structures. This fractality was quantified using the uncertainty exponent method, with uncertain points corresponding to the boundaries of the reentry basins. As fractality increases, so does the precision required in specifying initial conditions. Hence, from an operational point of view, it is easier to target the $5$-year reentry basin as opposed to the $25$-year basin. 
	This suggests that the enforcement of a lower mitigation guideline is,  from the dynamical perspective, advantageous for satellite mission design and disposal strategies.
\end{enumerate}

\section{Acknowledgments}
C.\,B.\, and Ch.\,S.\, acknowledge financial support from the National Research Foundation (NRF) of South Africa (grant number WABR240401211578). Ch.\,S.\, also acknowledges the hospitality of the Max Planck Institute for the Physics of Complex Systems in Dresden, Germany, during his visits in 2025 and 2026, when part of this work was conducted. The authors further thank the Centre for High Performance Computing in Cape Town (\url{https://www.chpc.ac.za}) and the University of Cape Town's High Performing Computing Facility (\url{https://www.ucthpc.uct.ac.za}) for providing the computational resources used in this study.


\newpage 
\appendix
	
	\section{Additional results in the $i-a$ space}\label{app:A}
	
	\subsection{Sensitivity of the maps with respect to the initial phases}
	\label{app:sub1}
	
\begin{figure}
		\centering
		\includegraphics[width=1\linewidth]{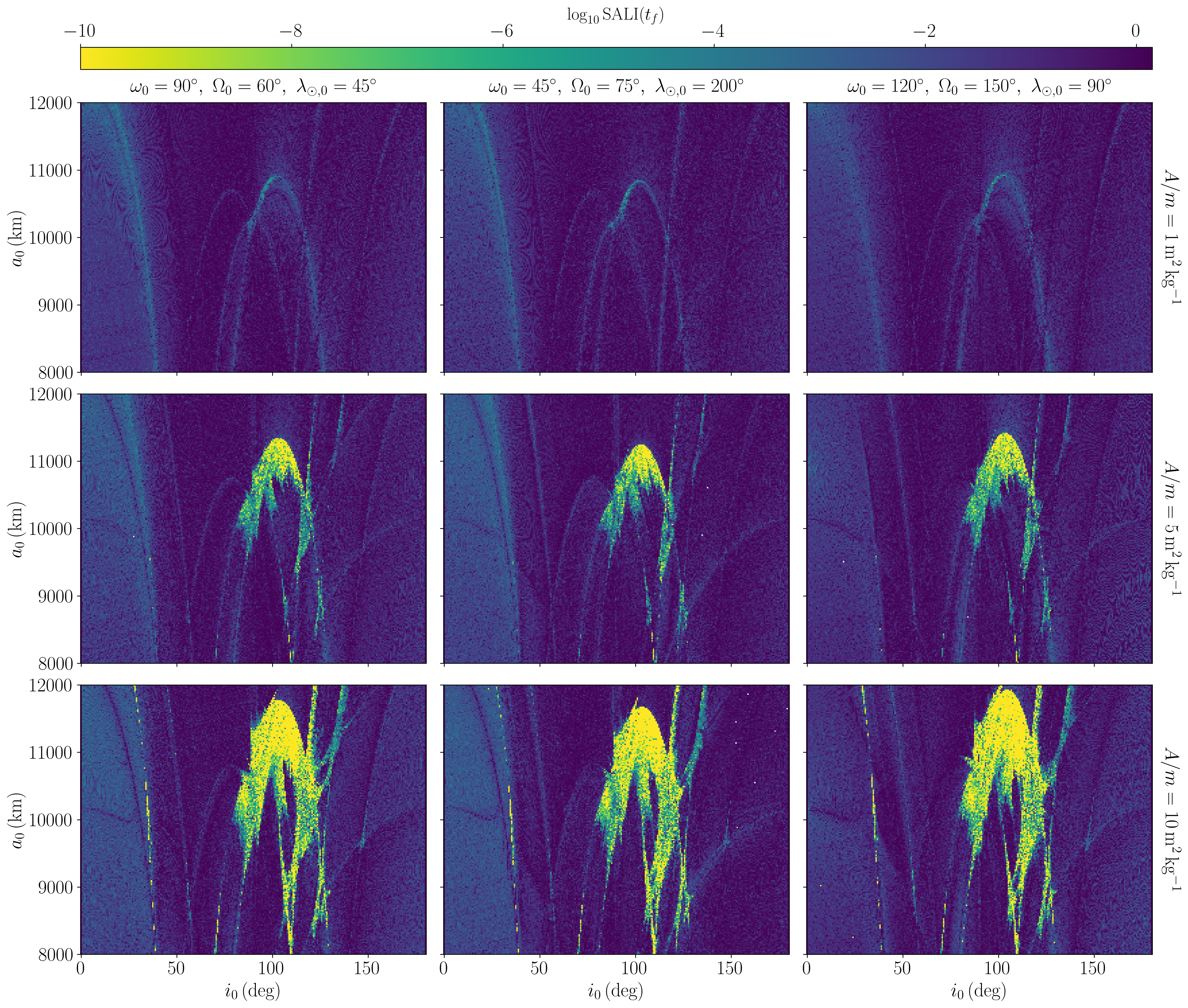}
		\caption{Dynamical maps when the phases of the system are varied. Initial conditions are colored according to the $\log_{10}\mbox{SALI}$ value  computed at $t_f=100$ years, over the interval $a_0\in[8\,000\mathrm{km},12\,000\mathrm{km}]$ and $i_0\in(0\degree,180\degree)$, with $e_0=0.05$ fixed and $\omega_0=\Omega_0=\lambda_{\odot,0}$ varied according to the column headings. Each row corresponds to $A/m\in\{1,5,10\}\,\mathrm{m}^2\mathrm{kg}^{-1}$. The observed dynamical structures do not significantly depend upon the phases of the system.}
		\label{fig:appfig1}
	\end{figure}

	Fig.\,\ref{fig:appfig1} shows several dynamical maps based on the SALI values at time $t_{f}=100$ years for $e_{0}=0.05$ when the initial phases of the system are varied, for $A/m\in\{1, 5, 10\}\,\mathrm{m}^2\mathrm{kg}^{-1}$. The dynamical structures (large chaotic seas or filamentary structures) are not significantly influenced by the initial phases of the system. Due to this lack of sensitivity, the initial phases can be set to any values without notable changes in the results, thereby justifying the choice made in the main text to set all initial phases to zero, namely $\omega_0=\Omega_0=\lambda_{\odot,0}=0\degree$.  
	
	\subsection{Additional maps when $e_{0}$ is varied}
	
\begin{figure}
		\centering
		\includegraphics[width=1\linewidth]{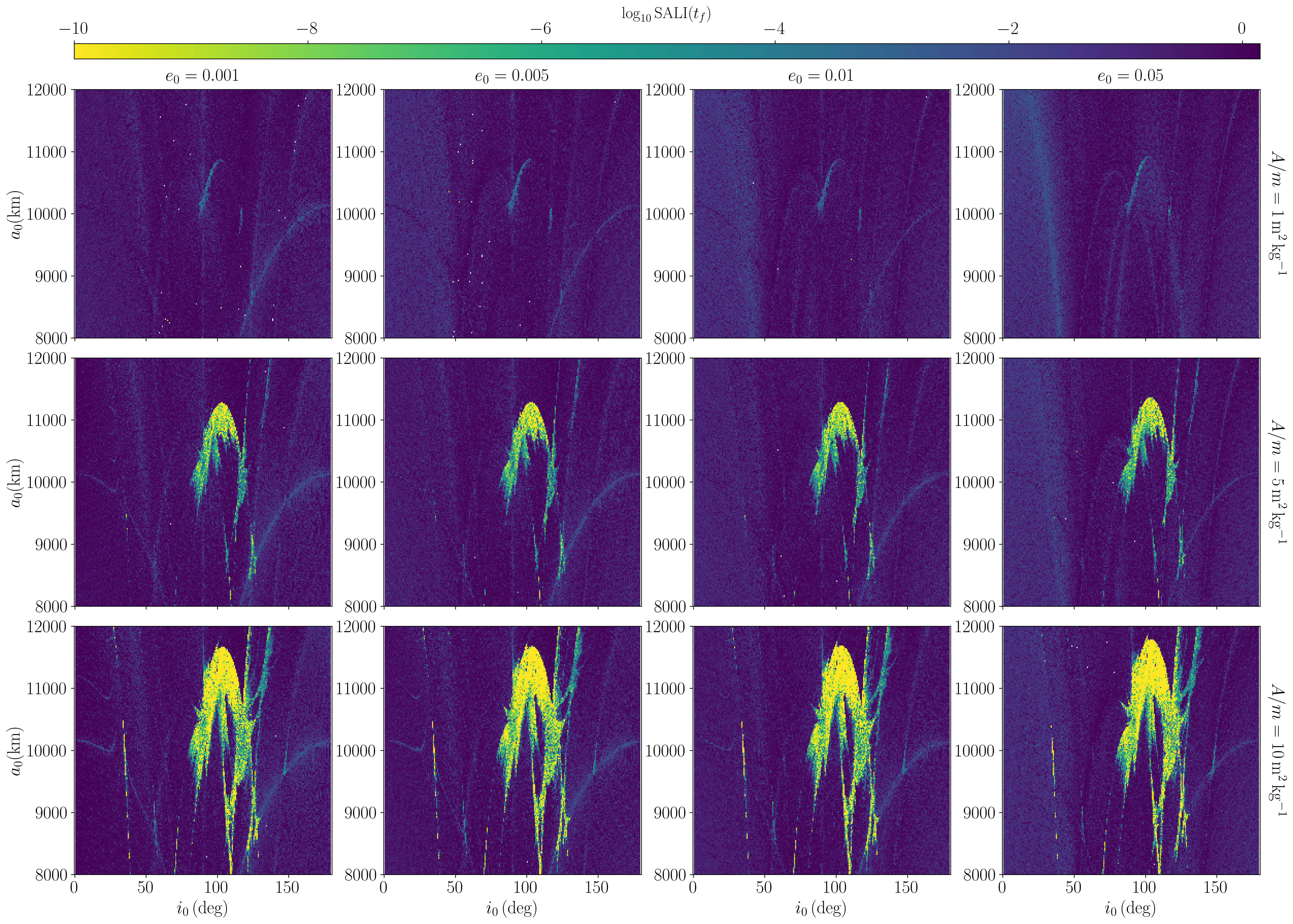}
		\caption{Dynamical maps when $e_{0}$ is varied. Initial conditions are colored according to the $\log_{10}\mbox{SALI}$ value  computed at $t_f=100$ years, over the interval $a_0\in[8\,000\mathrm{km},12\,000\mathrm{km}]$ and $i_0\in(0\degree,180\degree)$, with $\omega_0=\Omega_0=\lambda_{\odot,0}=0\degree$ fixed and $e_0\in\{0.001,0.005,0.01,0.05\}$ per column. Each row corresponds to $A/m\in\{1,5,10\}\,\mathrm{m}^2\mathrm{kg}^{-1}$. The observed dynamical structures of the system do not show a significant dependence on the value of $e_0$ within the considered range of initial eccentricities.
		}
		\label{fig:appfig2}
	\end{figure}

	In order to witness the changes (if any) to the chaotic structures present in the phase space when varying the value of $e_0$, Fig.\,\ref{fig:appfig2} shows several SALI maps computed at time $t_{f}=100$ years for $e_{0}\in\{0.001,0.005,0.01,0.05\}$. As in the case of the initial phases shown in Fig.\,\ref{fig:appfig1}, variations in  $e_0$ cause barely discernible differences in the chaotic seas and filaments across the corresponding maps, for all considered $A/m$ values, justifying the choice of $e_0=0.05$ throughout the main body of our investigation.
	
	\section{Stability and lifetime maps in the $i-e$ space}
	\label{app:B}
	
\begin{figure}
		\centering
		\includegraphics[width=1\linewidth]{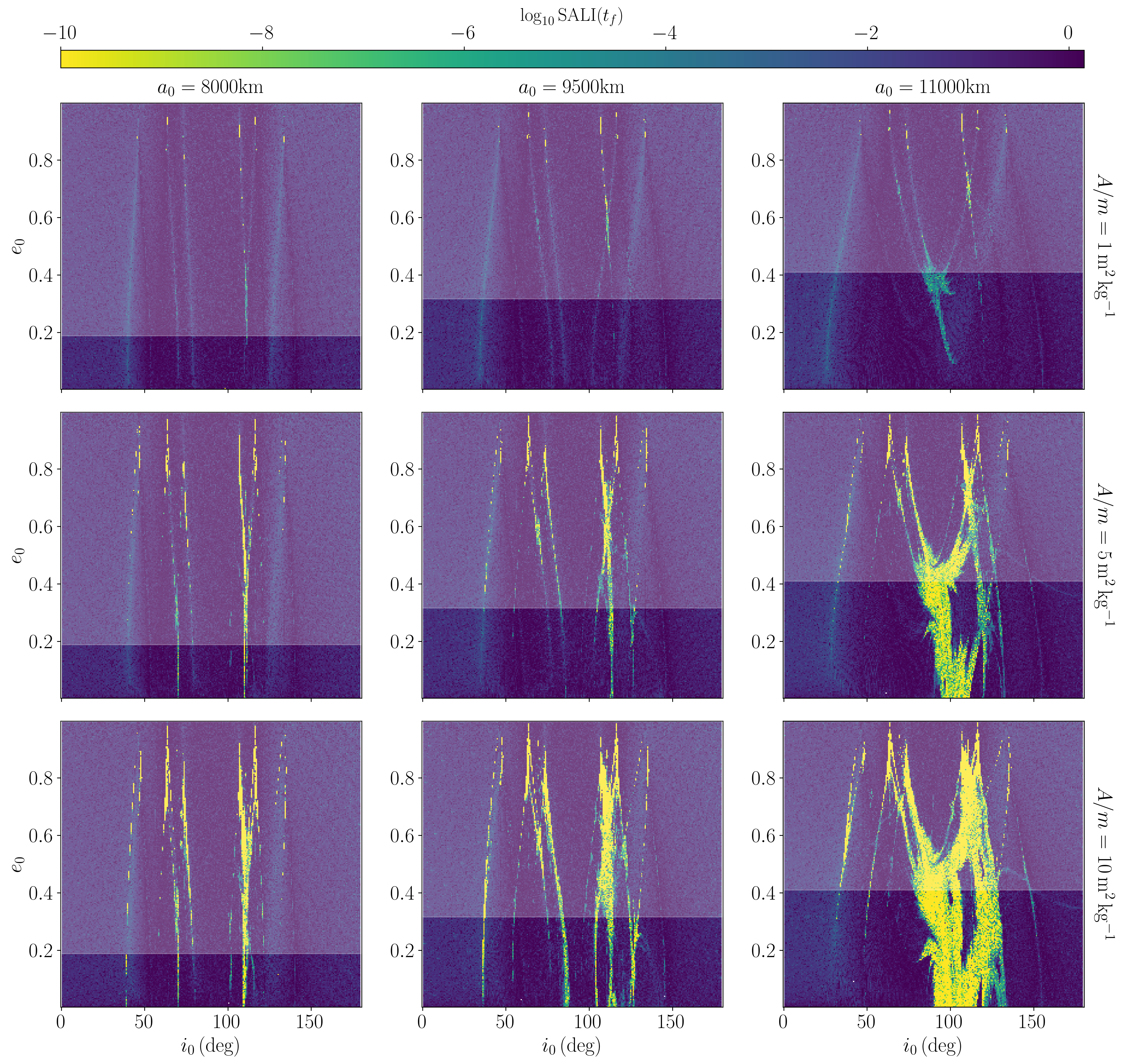}
		\caption{Initial conditions colored according to the $\log_{10}\mbox{SALI}$ value  computed at $t_f=100$ years, over the interval $i_0\in(0\degree, 180\degree)$ and $e_0\in(0,1)$, with $\omega_0=\Omega_0=\lambda_{\odot,0}=0\degree$ fixed and $a_0\in\{8\,000,9\,500,11\,000\}\,\mathrm{km}$ per column. Each row corresponds to $A/m\in\{1,5,10\}\,\mathrm{m}^2\mathrm{kg}^{-1}$. Nonphysical initial conditions are masked using a white sheen.}
		\label{fig:appfig3}
	\end{figure}

\begin{figure}
		\centering
		\includegraphics[width=1\linewidth]{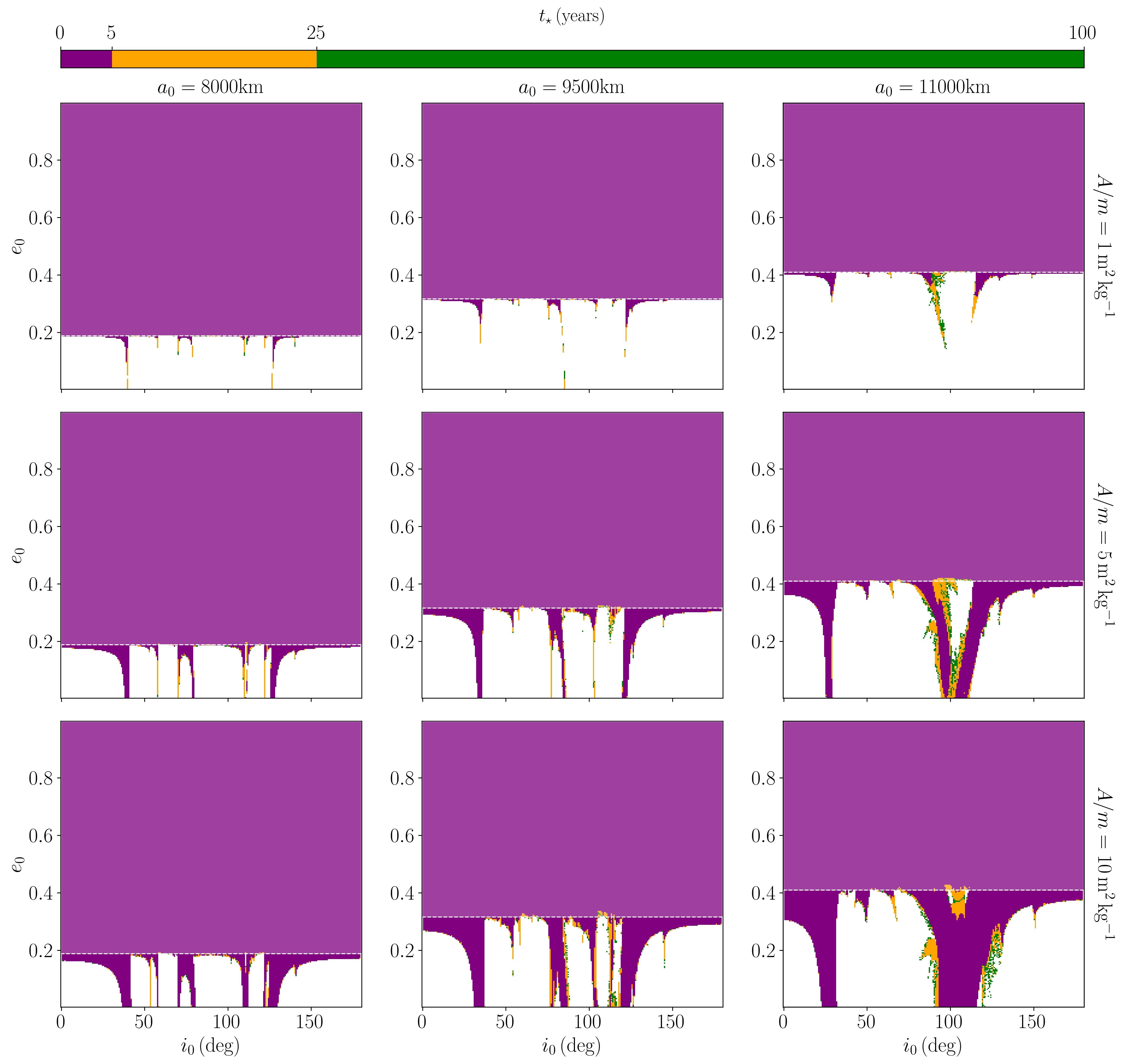}
		\caption{Initial conditions colored according to their orbital lifetime $t_\star$, over the interval $i_0\in(0\degree, 180\degree)$ and $e_0\in(0,1)$, with $\omega_0=\Omega_0=\lambda_{\odot,0}=0\degree$ fixed and $a_0\in\{8\,000,9\,500,11\,000\}\,\mathrm{km}$ per column. Each row corresponds to $A/m\in\{1,5,10\}\,\mathrm{m}^2\mathrm{kg}^{-1}$.}
		\label{fig:appfig4}
	\end{figure}

	To provide a more holistic visualization of the entire 6-dimensional phase space of the Hamiltonian in Eq.\,\eqref{eq:final_hamiltonian}, another slice of this phase space, namely $i-e$ space, is considered here. Mapping the SALI at $t_f=100$ years and the orbital lifetime respectively, Fig.\,\ref{fig:appfig3} and Fig.\,\ref{fig:appfig4} show initial conditions chosen over the range $i_0\in(0\degree, 180\degree)$ and $e_0\in(0,1)$, with $a_0$ taking on three different values, namely $a_0\in\{8\,000,9\,500,11\,000\}\,\mathrm{km}$, for $A/m\in\{1,5,10\}\,\mathrm{m}^2\mathrm{kg}^{-1}$. A white sheen is placed over a large upper portion of each map to identify  initial conditions corresponding to  an initial eccentricity $e_0$ exceeding the associated  critical eccentricity $e_\star$. These orbits are nonphysical; we plot them, however, for completeness' sake. As with the $i-a$ space in the main manuscript, the extent of chaos in the $i-e$ space increases with increasing $A/m$, with the largest amount of chaos observed for $a_0=11\,000\mathrm{km}$ and $A/m=10\,\mathrm{m}^2\mathrm{kg}^{-1}$. 
	
	Turning to the orbital lifetime maps of Fig.\,\ref{fig:appfig4}, while keeping in mind the SALI maps of Fig.\,\ref{fig:appfig3}, we see a correspondence (albeit not one-to-one, due to the broad band of  reentry orbits around $i_0\sim35\degree$ to $i_0\sim40\degree$) between the orbits that reenter within $25$ years and the chaotic regions revealed by the SALI. We see, in fact, that most orbits reenter within $5$ years, apart from the nonphysical orbits which naturally have $t_\star=0$.

	\section{Stability and lifetime maps in the $e-a$ space}  \label{app:C}
	
\begin{figure}
		\centering
		\includegraphics[width=1\linewidth]{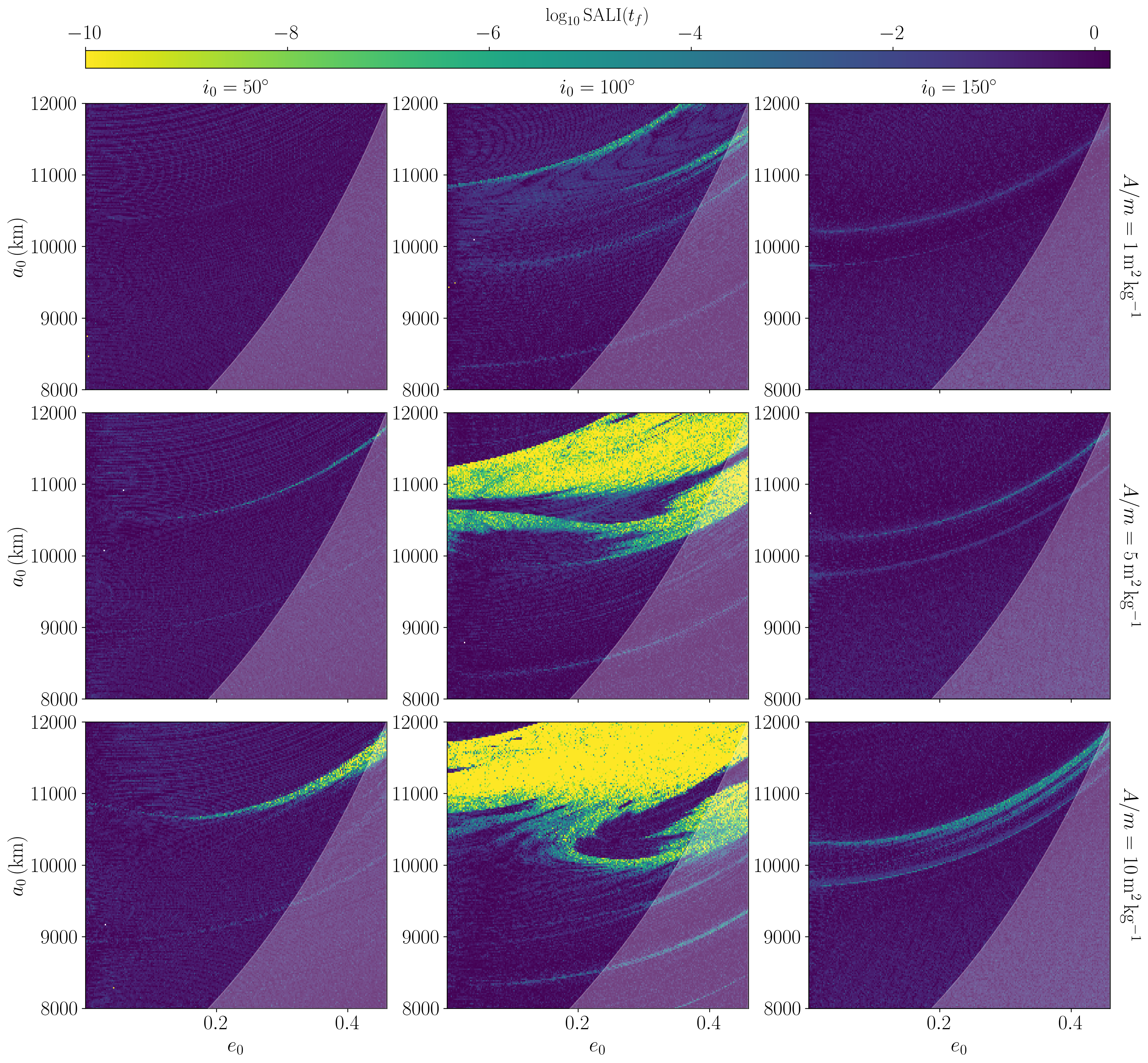}
		\caption{Initial conditions colored according to the $\log_{10}\mbox{SALI}$ value  computed at $t_f=100$ years, over the interval $e_0\in(0, 1)$ and $a_0\in[8\,000\mathrm{km},12\,000\mathrm{km}]$, with $\omega_0=\Omega_0=\lambda_{\odot,0}=0\degree$ fixed and $i_0\in\{50\degree,100\degree,150\degree\}$ per column. Each row corresponds to $A/m\in\{1,5,10\}\,\mathrm{m}^2\mathrm{kg}^{-1}$. Nonphysical initial conditions are masked using a white sheen.}
		\label{fig:appfig5}
	\end{figure}

\begin{figure}
		\centering
		\includegraphics[width=1\linewidth]{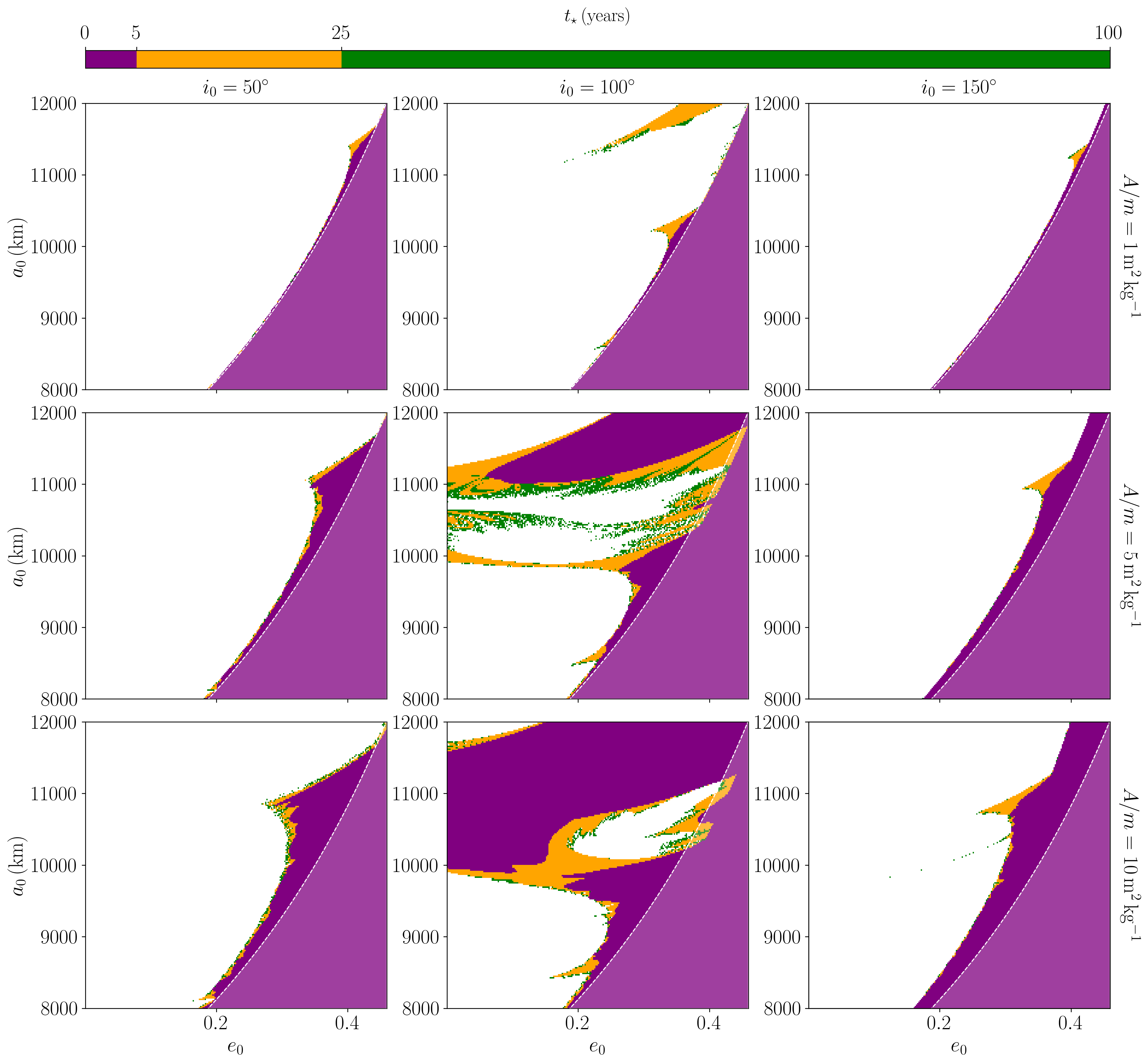}
		\caption{Initial conditions colored according to their orbital lifetime $t_\star$, over the interval $e_0\in(0, 1)$ and $a_0\in[8\,000\mathrm{km},12\,000\mathrm{km}]$, with $\omega_0=\Omega_0=\lambda_{\odot,0}=0\degree$ fixed and $i_0\in\{50\degree,100\degree,150\degree\}$ per column. Each row corresponds to $A/m\in\{1,5,10\}\,\mathrm{m}^2\mathrm{kg}^{-1}$.}
		\label{fig:appfig6}
	\end{figure}

	Continuing the exploration of the high-dimensional phase space of the Hamiltonian in Eq.\,\eqref{eq:final_hamiltonian}, Fig.\,\ref{fig:appfig5} and Fig.\,\ref{fig:appfig6} present maps of the SALI at $t_f=100$ years and the orbital lifetime, respectively, over initial conditions in the $e-a$ space. The scans are performed for  $e_0\in(0,1)$, $a_0\in[8\,000\mathrm{km},12\,000\mathrm{km}]$ and $i_0\in\{50\degree,100\degree,150\degree\}$, with $A/m\in\{1,5,10\}\,\mathrm{m}^2\mathrm{kg}^{-1}$. As before, nonphysical orbits are masked by a white sheen.
	
	The $e-a$ space again follows the trend of the $i-a$ and $i-e$ spaces, with the extent of chaotic regions increasing with $A/m$. The most pronounced chaotic behavior in this space occurs for $i_0\sim100\degree$, as expected from the prominent inverted U-shaped chaotic sea observed around $i_0\sim100\degree$ in 
	Fig.\,\ref{fig:fig2}. Furthermore, a qualitative correspondence can be identified between the chaotic regions in Fig.\,\ref{fig:appfig5} and the reentry structures in  Fig.\,\ref{fig:appfig6}. 
	
\newpage
\bibliographystyle{apalike}
\bibliography{biblio}

\end{document}